\documentclass[aps,prd,twocolumn]{revtex4-1}
\usepackage{graphicx}
\usepackage{textcomp}
\usepackage{amsmath}
\usepackage{amssymb}
\usepackage{amsthm}
\usepackage{braket}
\usepackage{fancyhdr}
\usepackage{slashed}
\usepackage{mathrsfs}
\usepackage{dsfont}
\usepackage{multirow} 
\usepackage{xcolor}

\begin{document}
\title{\Large Controlling Excited-State Contributions with Distillation in Lattice QCD Calculations of Nucleon Isovector Charges $g_S^{u-d}$, $g_A^{u-d}$, $g_T^{u-d}$}

\author{Colin\ Egerer}
\affiliation{Department of Physics, William and Mary, Williamsburg, VA 23185, USA}
\author{David Richards}
\author{Frank Winter}
\affiliation{Thomas Jefferson National Accelerator Facility, Newport News, VA 23606, USA}

\begin{abstract}
We investigate the application of the distillation smearing approach,
and the use of the variational method with an extended basis of
operators facilitated by this approach, on the calculation of the
nucleon isovector charges $g_S^{u-d}$, $g_A^{u-d}$, and $g_T^{u-d}$.  We find that the better
sampling of the lattice enabled through the use of distillation yields
a substantial reduction in the statistical uncertainty in comparison
with the use of alternative smearing methods, and furthermore,
appears to offer better control over the contribution of excited-states compared to use of a single, local interpolating operator.  The additional
benefit arising through the use of the variational method in the distillation approach is less
dramatic, but nevertheless significant given that it requires no additional Dirac inversions.
\end{abstract}
\maketitle

\section{Introduction}
The past decade has seen tremendous improvement in the ability of
Lattice QCD (LQCD) calculations to provide results that can confront
experiment. However, lattice computations of some key quantities
remain at odds with experimental determinations, including the
momentum fraction carried by partons in the nucleon, and, notably, the
axial-vector charge, $g_A^{u-d}$, of the nucleon. These discrepancies
are often attributed to finite-volume effects, and to the
contribution of excited states to ground-state matrix elements.  The
calculation of the axial-vector charge has been a particular focus
within the lattice community, and a dedicated effort to resolve these
discrepancies has demonstrated success\cite{Chang:2018uxx}, using a method to
control excited states inspired by the Feynman-Hellman theorem.  In
this paper we explore a different means of controlling excited states,
through the use of a novel smearing method,
``distillation''\cite{Peardon:2009gh}, and the variational method for
the case of the nucleon charges $g_A^{u-d}$, $g_S^{u-d}$ and $g_T^{u-d}$.

The matrix element of the flavor non-singlet axial-vector current
$A_\mu^a=\overline{\psi}\gamma_\mu\gamma_5\frac{\tau^a}{2}\psi$, where
$\psi$ is the isospin doublet of $u,d$ quark fields and $\tau$ the
Pauli-spin matrices acting in isospin space, between nucleons of
momenta $p$ and $p'$ can be expressed in terms of the axial-vector and the induced pseudoscalar and tensor form factors
\begin{align}
  \begin{split}
  &\bra{N\left(p',s'\right)}A_\mu^3\ket{N\left(p,s\right)}=\overline{u}_N\left(p',s'\right)\left[\gamma_\mu\gamma_5G_A\left(q^2\right)+\right. \\
    &\quad\left.+\frac{q_\mu}{2M_N}\gamma_5G_P\left(q^2\right)+i\frac{\sigma^{\mu\nu}q_\nu}{2M_N}\gamma_5G_T\left(q^2\right)\right]\frac{\tau^3}{2}u_N\left(p,s\right)
  \end{split}
  \label{eq:axial}
\end{align}
where $q_\mu=p_\mu'-p_\mu$ the momentum transfer. At zero momentum transfer, the contribution of the pseudoscalar and tensor form factors to the matrix element of equation \ref{eq:axial} vanish, and we are left with the definition of the isovector axial charge of the nucleon $g_A^{u-d}=G_A\left(0\right)$.
The axial-vector charge $g_A$ quantifies the $n\rightarrow
pe^{-}\overline{\nu}_e$ coupling, the degree of low-energy chiral
symmetry breaking in QCD, and even the difference between the $u$ and
$d$ quark contributions to the proton spin $g_A=\Delta u-\Delta d$.
The breadth of different phenomena dependent on a knowledge of $g_A$
highlights the need for precise agreement between theoretical and
experimental determinations. Moreover, the straightforward definition
of $g_A$ serves as a useful proving ground for any new lattice
algorithm purporting to calculate meaningful quantities in QCD.

Precise neutron $\beta$ decay experiments have measured the nucleon
isovector axial charge $g_A^{u-d}=1.2724\pm0.0023$ \cite{Tanabashi:2018oca}. Yet,
historically, the bulk of lattice calculations have systematically
under-determined $g_A^{u-d}$ by roughly 10-15\%. This has led to intense
discussion in the community on the role of finite-volume effects, the
lack of chiral symmetry in most lattice formulations of QCD, the
influence of heavy quark flavors, the use of local currents with large
discretization effects, and excited-state contamination might have on
the calculation of $g_A^{u-d}$ and nucleon matrix elements in
general. Calculations have been performed using the Highly Improved
Staggered Quark (HISQ) \cite{Lin:2018obj} and Domain Wall (DW) \cite{Ohta:2015aos}
fermion actions which, despite the restoration of chiral symmetry,
present challenges, namely, identifying quark flavors and the
numerical cost compared to Wilson-type actions, respectively. Calculations have
even been performed with $\mathcal{O}\left(a^2\right)$ improved
currents with novel noise reduction techniques
\cite{Liang:2016fgy,Yang:2015zja}, and others still with 2+1+1 flavor
QCD accounting for several excited-states in requisite fits
\cite{Gupta:2018qil,Gupta:2018wrq}. In all such cases the calculated value of
$g_A^{u-d}$ differs from experiment by $\sim5-10$\%. Clearly standard
techniques of calculating and subsequently fitting two- and three-point correlation
functions to extract $g_A^{u-d}$ routinely come up short.

Recent work by \cite{Chang:2018uxx} employs a methodology inspired by
the Feynman-Hellman theorem to control excited-state effects by
summing over all current insertion times, engendering extrapolation in
a single Euclidean temporal variable rather than two - agreement was
found to within 1\% of experiment. We remark that a few 2-flavor
lattice QCD calculations employing the Wilson plaquette/clover fermion
actions have been performed for $g_A^{u-d}$ whose results are within
$\sim1$\% of the experimental determination
\cite{Bali:2014nma,Horsley:2013ayv}. These studies calculated
$g_A^{u-d}$ for source-sink separations greater than 1 fm. Furthermore it
was found $g_A^{u-d}$ has a strong dependence on the spatial lattice
volume $V_3$. Taken together, suggesting the largest contributors to
uncertainty in $g_A^{u-d}$ lie in small lattice volumes and
excited-state contamination.

In this paper, we investigate a means of overcoming one of the
dominant systematic uncertainties in the calculation of nucleon
charges, namely that arising from our inability to isolate the
ground-state nucleon from its excitations. The variational method, an
algorithm to improve overlap onto a desired state, was applied in
\cite{Yoon:2016dij,Yoon:2016jzj,Owen:2012ts} to a small basis of
Jacobi-smeared nucleon interpolators of the form
$\chi\left(x\right)=\epsilon^{abc}\left[u^{a\top}\left(x\right)C\gamma_5d^b\left(x\right)\right]u^c\left(x\right)$,
where it was found variationally improved interpolators can reduce
excited-state effects. We will show that an amalgam of a suitable
basis of interpolating operators, the use of the variational method
with that basis, and an efficient means of computing the needed
correlation functions through the application of ``distillation'',
affords a powerful and computationally efficient method of taming
excited-state contributions to those charges. Furthermore, we will
show that ``distillation'' alone, enabling a momentum projection to be
performed at each time slice in the two- and three-point functions,
provides a significant increase in the statistical precision whilst
being an effective smearing operator with only a single interpolating
operator. This allows the reliable extraction of matrix elements at
earlier source-sink separations when the nucleon signal-to-noise ratio
is exponentially better. As lattice calculations proceed to ever more
complicated quantities, exemplified by quasi-PDFs and pseudo-PDFs, the
need for tamed excited-state effects is paramount.

In this work, we abstain from calculating and presenting renormalized
isovector charges of the ground-state nucleon, and from performing continuum,
finite-volume, and chiral extrapolations. Throughout this work we consider
forward-scattering matrix elements of nucleons at rest. Although much
of our motivating comments have centered around $g_A^{u-d}$, we extend
our technique to include the scalar $g_S^{u-d}$ and tensor $g_T^{u-d}$
charges of the nucleon, whose precise determination will constrain BSM
searches at the TeV scale and dark matter direct-detection searches. A
future study will explore forward-scattering matrix elements of moving
states, and generalizations to scalar, axial, and tensor nucleon form
factors.

This paper is organized as follows. In section~\ref{sec:method} we discuss
excited-state contamination and present the computational methodology
employed throughout this work. In particular, we review distillation
as a low-mode approximation of the more standard Jacobi and Wuppertal
smearing algorithms, and discuss the variational method as a means of
improving overlap onto a desired state. Section~\ref{sec:details}
describes the lattice ensemble, the explicit construction of the
interpolating fields, and then finishes with a discussion of the
strategy used to extract the nucleon charges from our data. In section~\ref{sec:results}
we first feature a comparison of the nucleon effective masses
obtained using a Jacobi-smeared interpolator, a single ``local''
distilled interpolator, and different variationally improved
interpolators from distinct bases of distilled interpolators. We then
proceed to present determinations of effective masses of the nucleon
via fits to our data and ultimately our extracted charges. We then conclude
with a discussion of our results, a cost-benefit analysis
of standard smearing techniques and distillation, implications for
assorted studies in nucleon structure, and directions for future
research.

\section{Computational Methodology}\label{sec:method}
We begin this section with some definitions we will use throughout
this work. Isovector charges $g_\Gamma$ of the nucleon are measured
experimentally through the neutron to proton transitions
$\bra{p\left(p,s\right)}\mathcal{O}_\Gamma^{ud}\ket{n\left(p,s\right)}$,
where $\mathcal{O}_\Gamma^{ud}=\overline{u}\Gamma d$, or via proton
and neutron charge differences. Imposing isospin symmetry, one can
show
\begin{equation}
  \bra{p\left(p,s\right)}\mathcal{O}_\Gamma^{ud}\ket{n\left(p,s\right)}=\bra{p\left(p,s\right)}\mathcal{O}_\Gamma^{u-d}\ket{p\left(p,s\right)}
  \label{eq:isospin}
\end{equation}
where $\mathcal{O}_\Gamma^{u-d}=\overline{u}\Gamma
u-\overline{d}\Gamma d$ is an external current. The isovector charges
of the nucleon are thus defined as
\begin{equation*}
  \bra{N\left(p,s\right)}\mathcal{O}_\Gamma^{u-d}\ket{N\left(p,s\right)}=g_\Gamma^{u-d}\overline{u}_s\left(p\right)\Gamma u_s\left(p\right)
\end{equation*}
where we normalize the nucleon spinors in Euclidean space according to
\begin{equation*}
  \sum_su_s\left(p\right)\overline{u}_s\left(p\right)=-i\slashed{p}_E+m_N.
\end{equation*}
Indeed the calculation of isovector quantities is computationally less
demanding than the calculation of isoscalar and flavor-diagonal
quantities, in which the cancellation of disconnected quark lines in
the isospin limit does not occur.

\subsection{Excited-State Contamination}
The calculation of hadronic matrix elements within LQCD requires the
construction of operators that attempt to interpolate a given hadronic
state from the vacuum, with minimal overlap with neighboring
states. As the precise wavefunction of any hadronic state is not
known, any operator construction of a desired continuum $J^{PC}$ is
merely a ``best guess'' and necessarily overlaps with other hadronic
states of the same quantum numbers - most notably, excited-states and
multi-particle states. This problem is compounded with the explicit
breaking of Lorentz symmetry, in which continuum operators residing in
different irreducible representations (irreps) of the Lorentz group
can mix under subduction to the same lattice irrep thereby increasing the
number of states contributing within a given lattice irrep.

Explicitly, consider a two-point correlation function projected to zero momentum
\begin{equation}
  C_{2\text{pt}}(t)=\sum_{\vec{x}}\langle\mathcal{O}(\vec{x},t)\overline{\mathcal{O}}(\vec{0},0)\rangle,
  \label{eq:simple_2pt}
\end{equation}
where $\mathcal{O}$ is an interpolating operator for the state of interest. Performing a spectral decomposition, this can be expressed as
\begin{equation}
  C_{2\text{pt}}(t)  = \sum_n\frac{1}{2E_n}\left|\bra{0}\mathcal{O}\ket{n}\right|^2e^{-E_nt},\label{eq:decomp}
\end{equation}
where the sum is over all states, of energy $E_n$, that can be created
with the operator $\mathcal{O}$. In order to extract reliably the
ground-state mass $M_0 \equiv E_0$, one can study the large time behavior of the two-point correlator, wherein contributions of excited states are negligible, or make a judicial
choice of interpolating operator such that the overlap factors $Z_n=\bra{0}\mathcal{O}\ket{n}$, for states $n>0$, are greatly suppressed relative to $\bra{0}\mathcal{O}\ket{n=0}$, and thereby determine $E_0$ at small
temporal separations.  Given that lattice calculations of baryon properties are constrained by an exponentially increasing noise-to-signal ratio with
increasing Euclidean time, the latter approach is far preferable. This issue is further compounded when considering
matrix elements of an external current $\mathcal{J}$, where the
suppression of excited-states is needed both between the source
interpolator and the current, and the current and the sink interpolator. Thus there is strong encouragement
to develop operators that couple predominantly to the ground-state, and only weakly to the excited-states.

\subsection{Smearing}
Interpolating fields constructed of point-like quark and gluonic
fields couple to hadronic states at all energy scales.
``Smearing'' is a well-established technique to increase the
overlap of interpolators with the low-lying states of the spectrum (i.e. confinement scale physics), and to reduce
the contribution of the high-energy modes to correlation functions, through the use of
spatially extended operators of hadronic size.  Specifically, we can
replace the quark fields $\psi(\vec{x},t)$ occurring in the path integral with spatially extended quark fields
\[
\tilde{\psi}(\vec{x},t) = \sum_{\vec{y}} S[U](\vec{x},\vec{y})
\psi(\vec{y},t),
\]
where $S[U](\vec{x},\vec{y})$ is a gauge-covariant scalar ``smearing''
kernel that is functionally dependent on the underlying gauge fields $U$ on some time slice $t$.

A frequently utilized smearing operator is $J_{\sigma,n_\sigma}\left(t\right)=\left(1+\frac{\sigma\nabla^2\left(t\right)}{n_\sigma}\right)^{n_\sigma}$ defining the Jacobi-smearing method \cite{Allton:1993wc},
where $\nabla^2$ is a gauge-covariant discretization of the Laplacian,
$\sigma$ is the smearing ``width'', and $n_\sigma$ represents the
number of applications of the Laplacian. In the large $n_\sigma$ limit, 
\begin{equation*}
  \tilde\psi\left(\vec{x},t\right)=\lim_{n_\sigma\rightarrow\infty}J_{\sigma,n_\sigma}\left(t\right)\psi\left(\vec{x},t\right)\rightarrow e^{\sigma\nabla^2\left(t\right)}\psi\left(\vec{x},t\right)
\end{equation*}
thus approaching a Gaussian of width $\sigma$, characteristic of the
size of a hadron.

\subsection{Distillation \& Operator Construction}
Distillation\cite{Peardon:2009gh} is a low-rank approximation to a gauge-covariant smearing kernel. Specializing to the
case of the Laplacian, we begin by seeking solutions to
$-\nabla^2\left(t\right)\xi^k\left(t\right)=\lambda^k\left(t\right)\xi^k\left(t\right)$. Ordering
solutions by the magnitude of the eigenvalues
$\lambda^k\left(t\right)$, the distillation operator is constructed as
the outer-product of two eigenvectors on a given time slice
\begin{equation}
  \square\left(\vec{x},\vec{y};t\right)_{ab}=\sum_{k=1}^N\xi_a^{\left(k\right)}\left(\vec{x},t\right)\xi_b^{\left(k\right)\dagger}\left(\vec{y},t\right),
\end{equation}
where the color indices $\lbrace a,b\rbrace$ are made explicit. The distillation operator
$\square$ is then applied to each quark or
antiquark field both at the source and at the sink.

Distillation affords several advantages over Jacobi-like smearing
methods.  Distillation, firstly, factorizes the construction of the
interpolating or current operators from quark propagation. Secondly,
distillation allows operators not only at the sink location but also
at the source to incorporate more elaborate spatial structure, and in
particular derivatives, without additional inversions of the Dirac
operator.  Finally, distillation enables momentum projection to be
performed both at the source and sink time slices, thus providing a
more thorough sampling of a given lattice configuration. In this case,
equation \ref{eq:simple_2pt} becomes
\begin{equation}
  C_{2\text{pt}}\left(t\right)=\sum_{\vec{x},\vec{y}}\langle\mathcal{O}\left(\vec{x},t\right)\overline{\mathcal{O}}\left(\vec{y},0\right)\rangle.\label{eq:dist2pt}
\end{equation}
These features have been key to
the precise determination of the many energy levels in the QCD spectrum
needed for the study of resonances in lattice calculations.  However,
these features are also of advantage in studies of hadron structure, by
facilitating more varied interpolator constructions, and by enabling momentum
projection at each time slice in a correlation function.

Specializing to the case of baryons, we will construct our
interpolating operators ${\cal O}_i$ following the procedure of
ref.~\cite{Edwards:2011jj,Dudek:2012ag}
\begin{equation}
  \mathcal{O}_i\left(t\right)\propto\epsilon^{abc}\mathcal{S}^{\alpha\beta\gamma}_i
  \left(\mathcal{D}_1\Box u\right)^\alpha_a\left(\mathcal{D}_2\Box d\right)^\beta_b\left(\mathcal{D}_3\Box u\right)^\gamma_c\left(t\right),
\end{equation}
where $\mathcal{D}_{1,2,3}$ are spatial operators
constructed from covariant derivatives, introduced to probe the
radial/angular structure of the nucleon wavefunction, and
$S^{\alpha\beta\gamma}$ are subduction matrices that project a state
of definite continuum spin into irreps of the hypercubic lattice (explicit construction given in Section \ref{subsec:distop_deets}). The building blocks of the two-point and three-point correlation functions
employing these interpolating operators are
\begin{itemize}
\item\textit{solution vectors}
  \begin{equation*}
    S^{\left(k\right)}_{\alpha\beta}(\vec{x},t';t)=M_{\alpha\beta}^{-1}\left(t',t\right)\xi^{(k)}\left(t\right)
  \end{equation*}
\item\textit{perambulators}
  \begin{equation*}
    \tau^{kl}_{\alpha\beta}\left(t',t\right)=\xi^{(k)\dagger}\left(t'\right)M^{-1}_{\alpha\beta}\left(t',t\right)\xi^{(l)}\left(t\right)
  \end{equation*}
\item\textit{elementals}
  \begin{align*}
    \Phi^{\left(i,j,k\right)}_{\alpha_1,\alpha_2,\alpha_3}\left(t\right)&=\epsilon^{abc}\left(\mathcal{D}_1\xi^{\left(i\right)}\right)^a\left(\mathcal{D}_2
    \xi^{\left(j\right)}\right)^b \\
    &\quad\left(\mathcal{D}_3 \xi^{\left(k\right)}\right)^c\left(t\right)S_{\alpha_1,\alpha_2,\alpha_3}
  \end{align*}
\end{itemize}
where it should be noted that the inversion of the lattice Dirac operator against a smeared point-source in standard techniques, is replaced in distillation by inversion of the lattice Dirac operator against the $k^{\text{th}}$ eigenvector of time slice $t$.
The principle disadvantage of the distillation method is that the number of
distillation eigenvectors $N$ needed to construct
a correlation function of the same resolution is expected to scale as
the lattice spatial volume $V_3$. Since the evaluation of the Wick contractions
for mesons and baryons scales as $N^3$ and $N^4$,
respectively, the volume-scaling cost is potentially severe.

\subsection{Variational Analysis}
\label{subsec:vari}
The variational method is an often employed technique in lattice
spectroscopy calculations that seeks to disentangle the contributions
of eigenstates to the two-point correlation of two interpolating
operators. The variational method begins by constructing a matrix of
correlation functions
\begin{equation}
  C_{ij}\left(t\right)=\langle\mathcal{O}_i\left(t\right)\overline{\mathcal{O}}_j\left(0\right)\rangle
  \label{eq:corrmatrix}
\end{equation}
where $\mathcal{O}_i$ and $\mathcal{O}_j$ belong to some basis $\mathcal{B}$ of appropriately constructed interpolators with identical quantum numbers. In practice, these interpolators transform with definite symmetries in quark flavor, derivative structure, and Dirac structure - where the Dirac structure of the operator is encoded in the subduction of a continuum operator into lattice irreps. The actual variational method begins by considering the system of generalized eigenvalue equations
\begin{equation}
  C\left(t\right)v_{\mathrm{\textbf{n}}}\left(t,t_0\right)=\lambda_{\mathrm{\textbf{n}}}\left(t,t_0\right)C\left(t_0\right)v_{\mathrm{\textbf{n}}}\left(t,t_0\right)
  \label{eq:gevp}
\end{equation}
with $\mathrm{\textbf{n}}\in\lbrace1,\cdots,\text{dim}\left(\mathcal{B}\right)\rbrace$ and $t>t_0$. It can be shown that at large times $t$ this system of equations is described by the ``principal correlators'' $\lambda_{\mathrm{\text{n}}}\left(t,t_0\right)=e^{-M_n\left(t-t_0\right)}$, where the trivial solution $v_{\mathrm{\text{n}}}=0$ is avoided by imposing the normalization condition $v_{\mathrm{\text{n'}}}^{\dagger}C\left(t_0\right)v_{\mathrm{\text{n}}}=\delta_{\mathrm{\text{n'n}}}$. Equation \ref{eq:gevp} is solved independently for each $t>t_0$, with $\lambda_{\mathrm{\text{n}}}\left(t_0,t_0\right)=1$ by construction. It is possible that in the process of solving Equation \ref{eq:gevp}, what appears to be the $m^{\text{th}}$ eigenvector on time slice $t_m$ is not the $m^{\text{th}}$ eigenvector on time slice $t_{m+1}$. To remedy this, we follow the procedure of \cite{Edwards:2011jj} by associating states on different time slices using the relative similarity between their associated eigenvectors. This is accomplished by selecting some reference time slice $t_{\text{ref}}$ and maximizing $v_{\mathrm{\text{n'}}}^{\text{ref}\dagger}C\left(t_0\right)v_{\mathrm{\text{n}}}$, where $v_{\mathrm{\text{n'}}}^{\text{ref}}\equiv v_{\mathrm{\text{n'}}}\left(t_{\text{ref}}\right)$.

A fitting function
\begin{equation}
  \lambda_{\mathrm{\text{n}}}\left(t,t_0\right)=\left(1-A_{\mathrm{\text{n}}}\right)e^{-M_{\mathrm{\text{n}}}\left(t-t_0\right)}+A_{\mathrm{\text{n}}}e^{-M'_{\mathrm{\text{n}}}\left(t-t_0\right)}
  \label{eq:princorfunc}
\end{equation}
is applied to the principal correlator to obtain the masses $M_{\mathrm{\text{n}}}$, $M_{\mathrm{\text{n}}}'$, and amplitude $A_{\mathrm{\text{n}}}$. The choice for $t_0$ is made by attempting to reconstruct the original correlation matrix $C_{ij}\left(t\right)$ using the masses $M_{\mathrm{\text{n}}}$, $M_{\mathrm{\text{n}}}'$ and the extracted overlap factors $Z_i^{\mathrm{\text{n}}}$ \cite{Dudek:2007wv}. The degree of agreement for $t>t_0$ dictates the choice for $t_0$. Solving the variational method introduces a slight time-dependence in the overlap factors. The overlap factors are thus evaluated on a time slice $t_Z>t_0$ that minimizes differences between original/reconstructed correlators \cite{Dudek:2007wv}. The resulting eigenvectors $v_{\mathrm{\textbf{n}}}$ for each $t>t_0$ yield the optimal linear combination of $\mathcal{O}_i\in\mathcal{B}$ to project the $\mathrm{\textbf{n}}^{\text{th}}$-state from the vacuum
\begin{equation}
  \mathcal{O}^{\text{opt}}_{\textbf{n}}=\sum_iv_{\mathrm{\textbf{n}}}^i\mathcal{O}_i.
\end{equation}
The variational method may equally be applied to correlation matrices comprised of three-point functions of different interpolating fields. This would necessarily produce a better determined $\mathcal{O}^{\text{opt}}_{\textbf{n}}$, but due to the high Wick contraction cost of using distillation we elected to perform the variational method on a correlation matrix of two-point functions, thereby fixing $\mathcal{O}^{\text{opt}}_{\textbf{n}}$ for later use.

\section{Computational Details}\label{sec:details}
Our analysis considers a 350 configuration ensemble of 2+1 flavor QCD
using the clover-Wilson fermion action, where the associated gauge
links are smeared by one application of the stout smearing
\cite{Morningstar:2003gk} algorithm. This smearing yields a
tadpole-improved tree-level clover coefficient that is nearly
identical to the corresponding non-perturbative determination. The
reader is referred to \cite{Yoon:2016dij,Yoon:2016jzj} for a
discussion regarding the gauge action used to generate this
ensemble. Calculations were performed on a $32^3\times64$ lattice with
periodic (spatial) and anti-periodic (temporal) boundary conditions
and an inverse coupling of $\beta=6.3$. In the three flavor theory,
the strange quark mass was set by requiring the ratio
$\left(2M_{K^+}^2-M_{\pi^+}^2\right)/M_{\Omega^-}$ to assume its
physical value.  This lattice ensemble was found to have a lattice
spacing of $a=0.09840(4)\text{ fm}$ via the Wilson-flow scale $w_0$ ,
and with $am_\pi\sim0.176803$ yielding $am_{\pi}L\sim5.658$ and $m_\pi
= 356~{\rm MeV}$.

We explore the efficacy of four different types of operators used to
interpolate ground-state nucleons from the vacuum, with particular
consideration given to distillation. In this work we only study
zero-momentum nucleons, polarized along the z-axis. A future work will
study nucleons with non-zero momentum.

\subsubsection*{Jacobi smeared Interpolator}
Prior to construction of our quark sources comprising our selected
interpolators, the background gauge links are smeared via 20
applications of stout smearing with smearing parameter:
$\rho_{ij}=0.08$ and $\rho_{\mu4}=\rho_{4\mu}=0$, where
$\rho_{\mu\nu}$ quantifies the weight given to staple links aligned in
the $\mu\nu$-plane when constructing the smeared links. Such gauge
smearing is essential for reducing noise present in the resultant
correlation functions due to source fluctuations. Before inverting the
Dirac operator against the smeared sources, we apply a single
interation of stout smearing with $\rho=0.125$ to the gauge links -
thereby avoiding potentially small eigenvalues in the inversion.

As a benchmark with which to compare to distillation, we begin with the simplest nucleon interpolator consistent with the nucleon $J^{PC}$
\begin{equation}
  \mathcal{N}\left(x\right)=\epsilon^{abc}\left[u^{a\top}\left(x\right)C\frac{\left(1\pm\gamma_4\right)}{2}\gamma_5 d^b\left(x\right)\right]u^c\left(x\right)
  \label{eq:standardNuc}
\end{equation}
where $u,d$ are the two flavors of (degenerate) light quarks, $\lbrace a,b,c\rbrace$ color indices, $C=\gamma_2\gamma_4$ the charge conjugation matrix, and a suppressed Dirac index. The non-relativistic projector $\left(1\pm\gamma_4\right)/2$ is included in the operator construction to reduce the noise-to-signal ratio in forward (backward) propagating states. To make contact with previous work (e.g. \cite{Yoon:2016dij,Yoon:2016jzj}) $\mathcal{N}\left(x\right)$ is smeared with 60 hits of Jacobi smearing, with width $\sigma=5.0$. We refer the reader to \cite{Yoon:2016dij,Yoon:2016jzj} for an extensive analysis of the effect different Jacobi smearing parameters has on the determination of nucleon isovector charges. Herein we refer to the Jacobi smeared interpolator as ``Jacobi-SS''.

Correlation functions that employ $\mathcal{N}$ as the source/sink interpolator are constructed via application of appropriate projection operators
\begin{equation}
  C^{2\text{pt}}\left(t\right)=\sum_{\vec{x}}\langle\mathcal{P}^{2\text{pt}}_{\beta\alpha}\mathcal{N}_\alpha\left(\vec{x},t\right)\overline{\mathcal{N}}_\beta\left(0\right)\rangle
\end{equation}
\begin{equation}
  C^{3\text{pt}}\left(t,\tau\right)=\sum_{\vec{x},\vec{z}}\langle\mathcal{P}^{3\text{pt}}_{\beta\alpha}\mathcal{N}_\alpha\left(\vec{x},t\right)\mathcal{O}_\Gamma^{u-d}\left(\vec{z},\tau\right)\overline{\mathcal{N}}_\beta\left(0\right)\rangle
\end{equation}
where $\mathcal{P}^{2\text{pt}}=\left(1+\gamma_4\right)/2$ is used to project onto the forward-propagating positive-parity nucleon, and where $\mathcal{P}^{3\text{pt}}=\mathcal{P}^{2\text{pt}}\left(1+i\gamma_5\gamma_3\right)$ is used for the corresponding connected insertions. The spatial sums serve to project each correlation function to zero momentum. A standard spectral decomposition demonstrates that the Dirac structure of $\mathcal{O}_\Gamma^{u-d}$ must be $1$, $\gamma_4$, $\gamma_3\gamma_5$, $\gamma_1\gamma_2$ to extract the scalar, vector, axial, and tensor charges, respectively. Lastly, the sequential source method is implemented to calculate $C^{3\text{pt}}$, thereby minimizing the number of distinct inversions of the Dirac operator.

\subsubsection*{Distilled Interpolators}
\label{subsec:distop_deets}
We use a distillation space of rank 64, from which the
perambulators and solution vectors are constructed. The distillation
space on each time slice is calculated only after 10 iterations of
stout smearing is applied to the gauge links with smearing factor
$\rho_{ij}=0.08$ and $\rho_{\mu4}=\rho_{4\mu}=0$.

When expressed in a form exposing the permutational symmetry of the flavor ($\mathcal{F}$), spatial ($\mathcal{D}$) and Dirac ($\mathcal{S}$) structures, our distilled interpolators take the form
\begin{equation}
  \mathcal{O}=\left(\mathcal{F}_{\mathcal{P}\left(\text{F}\right)}\otimes\mathcal{S}_{\mathcal{P}\left(\text{S}\right)}\otimes\mathcal{D}_{\mathcal{P}\left(\text{D}\right)}\right)\lbrace q_1q_2q_3\rbrace
\end{equation}
where $\mathcal{P}\left(\cdots\right)$ expresses the symmetric (S), mixed-symmetric (M), and anti-symmetric (A) character of the given structure. Explicitly our employed distilled interpolators are
\begin{itemize}
  \centering
\item \resizebox{0.3\textwidth}{!}{$\left(N_M\otimes\left(\frac{1}{2}^+\right)^1_M\otimes D^{[0]}_{L=0,S}\right)^{J^P=\frac{1}{2}^+}=\thinspace^2S_S\tfrac{1}{2}^+$}
\item \resizebox{0.3\textwidth}{!}{$\left(N_M\otimes\left(\frac{1}{2}^+\right)^1_M\otimes D^{[2]}_{L=0,M}\right)^{J^P=\frac{1}{2}^+}=\thinspace^2S_M\tfrac{1}{2}^+$}
\item \resizebox{0.3\textwidth}{!}{$\left(N_M\otimes\left(\frac{1}{2}^+\right)^1_M\otimes D^{[2]}_{L=0,S}\right)^{J^P=\frac{1}{2}^+}=\thinspace^2S'_S\tfrac{1}{2}^+$}
\item \resizebox{0.3\textwidth}{!}{$\left(N_M\otimes\left(\frac{1}{2}^+\right)^1_M\otimes D^{[2]}_{L=1,A}\right)^{J^P=\frac{1}{2}^+}=\thinspace^2P_A\tfrac{1}{2}^+$}
\item \resizebox{0.3\textwidth}{!}{$\left(N_M\otimes\left(\frac{1}{2}^+\right)^1_M\otimes D^{[2]}_{L=1,M}\right)^{J^P=\frac{1}{2}^+}=\thinspace^2P_M\tfrac{1}{2}^+$}
\item \resizebox{0.3\textwidth}{!}{$\left(N_M\otimes\left(\frac{3}{2}^+\right)^1_S\otimes D^{[2]}_{L=1,M}\right)^{J^P=\frac{1}{2}^+}=\thinspace^4P_M\tfrac{1}{2}^+$}
\item \resizebox{0.3\textwidth}{!}{$\left(N_M\otimes\left(\frac{3}{2}^+\right)^1_S\otimes D^{[2]}_{L=2,M}\right)^{J^P=\frac{1}{2}^+}=\thinspace^4D_M\tfrac{1}{2}^+$}
\end{itemize}

\noindent where the superscript $J^P$ indicates the overall spin-parity quantum numbers of the interpolator \cite{Edwards:2011jj}. For brevity, we have expressed the interpolators in a compact spectroscopic notation $^{2S+1}L_\mathcal{P}J^{P}$, where $S$ represents the Dirac spin, $L$ the angular momentum induced by any derivatives, $\mathcal{P}$ the permutational symmetry of the derivatives, and $J^P$ the total angular momentum and parity of the interpolator. The first distilled interpolator we consider is the $^2S_S\frac{1}{2}^+$ interpolator, which is the closest non-relativistic analogue of the standard nucleon interpolator given in Eq. \ref{eq:standardNuc}.

Our first application of the variational method is to a basis of three distilled interpolators
\begin{equation}
  \mathcal{B}_3=\lbrace\thinspace^2S_S\tfrac{1}{2}^+,\thinspace^4P_M\tfrac{1}{2}^+,\thinspace^4D_M\tfrac{1}{2}^+\rbrace
\end{equation}
where we note $\thinspace^4P_M\tfrac{1}{2}^+$ and $\thinspace^4D_M\tfrac{1}{2}^+$ are explicitly of hybrid character. This choice is principally motivated by \cite{Dudek:2012ag} where it was found these hybrid interpolators, in addition to $\thinspace^2S_S\tfrac{1}{2}^+$, had predominant overlap onto the ground-state nucleon. The variational method applied to $\mathcal{B}_3$ leads to a variationally improved interpolator that we define to be $\hat{\mathcal{P}}_3$. The final interpolator we consider is found by expanding the basis $\mathcal{B}_3$ to include distilled interpolators that probe the radial/orbital structure of the nucleon (see above)
\begin{align}
  \begin{split}
    \mathcal{B}_7&=\lbrace\thinspace^2S_S\tfrac{1}{2}^+,\thinspace^2S_M\tfrac{1}{2}^+,\thinspace^2S'_S\tfrac{1}{2}^+,\thinspace^2P_A\tfrac{1}{2}^+,\\
    &\qquad\qquad \thinspace^2P_M\tfrac{1}{2}^+,\thinspace^4P_M\tfrac{1}{2}^+,\thinspace^4D_M\tfrac{1}{2}^+\rbrace.
  \end{split}
\end{align}
We refer to this variationally improved interpolator as $\hat{\mathcal{P}}_7$. The superficially redundant $\thinspace^2S_S\tfrac{1}{2}^+$ and $\thinspace^2S'_S\tfrac{1}{2}^+$ interpolators, with the same spectroscopic notation but differing derivative constructions, correspond to different radial extents of the interpolator.

As outlined in Section \ref{subsec:vari}, the construction of a variationally-improved interpolator relies on careful selection of a reference time slice $t_0$, and the time slice $t_Z$ at which to evaluate the overlap factors. Dividing out the ground-state time dependence, if a single exponential were to contribute to the principal correlator, a plateau of unity should be expected. Based on the applied fits of equation \ref{eq:princorfunc}, a good determination of the ground-state within our basis of interpolators is thus indicated by a plateau in the re-scaled correlator around unity - $A_0\ll1$ and $\Delta m_{n'0}=m_{n'}-m_0\gg1$. For the two variationally-improved interpolators we consider, we found
\begin{align*}
  \hat{\mathcal{P}}_3&\longrightarrow\lbrace t_0=3, t_Z=5\rbrace \\
  \hat{\mathcal{P}}_7&\longrightarrow\lbrace t_0=6, t_Z=8\rbrace
\end{align*}
led to ideal reconstruction of the original correlator.

\subsection{Matrix Element Extraction}
The extraction of matrix elements on the lattice typically proceeds by
calculation of some 3-point correlator, over a range of source-sink
interpolator separations, and where an external current is inserted
for all intermediate times. Under the presumption of no excited-state
contamination (i.e. in the limits $0\ll\tau\ll t_{\text{sep}}$), the
3-point correlator is then divided per ensemble average by a two point
correlator with the same source/sink interpolators of the same
source-sink separation. This division removes overlap factors, masses,
and exponential source/sink time dependence from the matrix element
signal. A plateau in this ratio, for fixed interpolator separations
and varying current insertion times, should be the desired matrix
element. However, as we are interested in ground-state matrix elements
$\mathcal{J}_{00}$, the plateau necessarily contains contributions
from matrix elements of all excited-states.

\subsubsection{Correlator Behavior}
Our two-point correlation function using Jacobi-smeared interpolators is defined by
\begin{equation*}
  C^{2\text{pt}}_{\alpha\beta}(t)=\sum_{\vec{x}}\langle\mathcal{N}_\alpha\left(\vec{x},t\right)\overline{\mathcal{N}}_\beta(\vec{0},0)\rangle.
\end{equation*}
Inserting the non-relativistic projector, and performing a spectral decomposition exposes the competing contributions of all states in the spectrum:
\begin{equation}
  C^{2\text{pt}}\left(t\right)=2\sum_n\left|Z_n\right|^2e^{-M_nt},
  \label{eq:2ptsimple}
\end{equation}
where the sum is only over eigenstates with quantum numbers of $\mathcal{N}$. To quantify and control excited-state contributions to $C^{2\text{pt}}$, we elect to perform a 2-state fit of the form
\begin{equation}
  C^{\text{2pt}}_{\text{fit}}\left(t\right)=e^{-M_0t}\left[\left|\mathbf{a}\right|^2+\left|\mathbf{b}\right|^2e^{-\left(M_1-M_0\right)t}\right]
  \label{eq:2ptfit}
\end{equation}
for each of the four distinct interpolators we consider. The factoring
of the ground-state time dependence aids in stabilizing our fits and
in the determination of our extracted charges, as explained later. In
this manner we obtain determinations of $M_0$, $M_1$,
$\left|Z_0\right|^2$ and $\left|Z_1\right|^2$, while simultaneously
quantifying the efficacy of each interpolator to separate the
ground-state from its excitations \textit{viz}.\ $\Delta
m=M_1-M_0$. The correlator behavior when using distilled interpolators
is identical to that above, except for the addition of an overall
volume factor $V_3$ due to the momentum projection at the source
implied by eqn.~\ref{eq:dist2pt}.

A zero momentum projected three-point correlation function using Jacobi-smeared interpolators is given by
\begin{equation}
  C_{\alpha\beta}^{3\text{pt}}(t_{\text{sep}};\tau)=\sum_{\vec{x}}\sum_{\vec{z}}\langle\mathcal{N}_\alpha\left(\vec{x},t_{\text{sep}}\right)\mathcal{J}\left(\vec{z},\tau\right)\overline{\mathcal{N}}_\beta(\vec{0},0)\rangle,
  \label{eq:3ptsimple}
\end{equation}
where $\tau$ is the insertion time slice, restricted to the time-ordering $0<\tau<t_{\text{sep}}$, and $\mathcal{J}$ the external current. An analogous spectral decomposition with the appropriate projector leads to
\begin{align}
  \begin{split}
  C^{3\text{pt}}&(t_{\text{sep}};\tau)=\sum_{n,s}\sum_{n',s'}\frac{e^{-M_{n'}\left(t_{\text{sep}}-\tau\right)}e^{-M_n\tau}}{4M_{n'}M_n}\times \\
  &\quad\quad\times Z_{n'}Z_n^\dagger\bra{n',p',s'}\mathcal{J}\ket{n,p,s}.
  \end{split}
  \label{eq:3ptij}
\end{align}
Again retaining the lowest two energy eigenstates in the sum, we have
\begin{align*}
  C^{3\text{pt}}&\left(t_{\text{sep}},\tau\right)=\left(\frac{\left|Z_0\right|^2}{4M_0^2}\mathcal{J}_{00}e^{-M_0t_{\text{sep}}}+\frac{\left|Z_1\right|^2}{4M_1^2}\mathcal{J}_{11}e^{-M_1t_{\text{sep}}}\right)+ \\
  &+\left(\frac{Z_0Z_1^\dagger}{4M_0M_1}\mathcal{J}_{01}e^{-M_0t_{\text{sep}}}e^{-\left(M_1-M_0\right)\tau}+\right. \\
  &\left.\quad\quad\quad\quad\quad\quad+\frac{Z_1Z_0^\dagger}{4M_0M_1}\mathcal{J}_{10}e^{-M_1t_{\text{sep}}}e^{\left(M_1-M_0\right)\tau}\right)
\end{align*}
where $\mathcal{J}_{n'n}=\bra{n',s'}\mathcal{J}\ket{n,s}$, with $n',n\in\mathbb{N}$. By isolating the current insertion time dependence of the three-point correlator, it becomes clear that calculation of a three-point correlation function for a single source-sink separation $t_{\text{sep}}$ is insufficient to reliably extract the matrix elements $\mathcal{J}_{00}$ and $\mathcal{J}_{11}$.
As $Z_{\mathrm{\text{n}}}$ are real and $\mathcal{J}_{01}=\mathcal{J}_{10}$ for zero-momentum states, the above can be reorganized into
\begin{align*}
  \begin{split}
  C^{\text{3pt}}\left(t_{\text{sep}},\tau\right)&=\left(\frac{\left|Z_0\right|^2}{4M_0^2}\mathcal{J}_{00}e^{-M_0t_{\text{sep}}}+\frac{\left|Z_1\right|^2}{4M_1^2}\mathcal{J}_{11}e^{-M_1t_{\text{sep}}}\right) \\
  &+\frac{Z_0Z_1}{2M_0M_1}\mathcal{J}_{01}e^{-\frac{\left(M_1+M_0\right)}{2}t_{\text{sep}}}\times \\
  &\qquad\qquad\times\cosh\left[\left(M_1-M_0\right)\left(\tau-\tfrac{t_{\text{sep}}}{2}\right)\right].
  \end{split}
\end{align*}
We then apply a 2-state fit to the three-point correlation functions,
\begin{align}
  \begin{split}
    C^{3\text{pt}}_{\text{fit}}&\left(t_{\text{sep}},\tau\right)=e^{-M_0t_{\text{sep}}}\left(\mathcal{A}+\mathcal{B}e^{-\Delta mt_{\text{sep}}}+\right. \\
    &\left.\qquad+\thinspace\mathcal{C}e^{-\Delta m\frac{t_{\text{sep}}}{2}}\cosh\left[\Delta m\left(\tau-\tfrac{t_{\text{sep}}}{2}\right)\right]\right)
  \end{split}
  \label{eq:3ptfit}
\end{align}
where $\tau$ is the current insertion time and $\Delta m=M_1-M_0$. We
note that we retain $M_0$ and $M_1$ as parameters in our fit, rather
than the difference $\Delta m$. As with the functional form employed
to fit the two-point correlation functions, we factor the ground-state
time dependence from the functional form of the three-point fits. This
factoring makes manifest the desired ground-state matrix element in
the limits $0\ll\tau\ll t_{\text{sep}}$. The correlator behavior when
using distilled interpolators is again identical to that above, except
for the addition of an overall volume factor $V_3$ to Equation
\ref{eq:3ptsimple}.

Determining the precise relationship between the fitted parameters to extract the ground-state matrix element $\mathcal{J}_{00}$, requires a more detailed look at our interpolators. Although the constructed distilled interpolators contain no free spinor indices, our use of positive-parity nucleons polarized along the $z$-direction can be viewed as the standard application of projectors on the nucleon interpolating field of Equation \ref{eq:standardNuc}. At zero-momentum we again have,
\begin{align*}
  C^{\text{2pt}}&\left(t\right)=\sum_{\vec{x}}\langle\mathcal{P}^{\text{2pt}}_{\beta\alpha}\mathcal{N}_\alpha\left(\vec{x},t\right)\overline{\mathcal{N}}_\beta\left(0\right)\rangle \\
  C^{\text{3pt}}\left(t_{\text{sep}},\tau\right)&=\sum_{\vec{x},\vec{z}}\langle\mathcal{P}^{\text{3pt}}_{\beta\alpha}\mathcal{N}_\alpha\left(\vec{x},t_{\text{sep}}\right)\mathcal{O}_\Gamma^{u-d}\left(\vec{z},\tau\right)\overline{\mathcal{N}}_\beta\left(0\right)\rangle.
\end{align*}
Proceeding with the spectral decomposition of $C^{2\text{pt}}$ we have,
\begin{align*}
  C^{2\text{pt}}&\left(t_{\text{sep}}\right)=\sum_{n,s}\frac{e^{-M_nt_{\text{sep}}}}{2M_n}\mathcal{P}^{\text{2pt}}_{\beta\alpha}\times \\
  &\qquad\qquad\qquad\times\bra{\Omega}\mathcal{N}_\alpha\ket{n,p,s}\bra{n,p,s}\mathcal{N}_\beta^\dagger\ket{\Omega} \\
  &=\sum_{n,s}\frac{\left|Z_n\right|^2e^{-M_nt_{\text{sep}}}}{2M_n}\mathcal{P}^{\text{2pt}}_{\beta\alpha}u^n_\alpha(\vec{0},s)\overline{u}^n_\beta(\vec{0},s) \\
  &=\sum_n\frac{\left|Z_n\right|^2e^{-M_nt_{\text{sep}}}}{2M_n}\text{Tr}\left[\mathcal{P}^{\text{2pt}}\left(-i\slashed{p}_E+M_n\right)\right].
\end{align*}
From which it is easily shown that the constant coefficients of the 2-state fit applied our two-point correlators are of the form $\left|\mathbf{a}\right|^2=2\left|Z_0\right|^2$ and $\left|\mathbf{b}\right|^2=2\left|Z_1\right|^2$.

The spectral decomposition of $C^{3\text{pt}}$ proceeds analogously,
\begin{align*}
  C^{3\text{pt}}&\left(t_{\text{sep}},\tau\right)=\sum_{n,s}\sum_{n',s'}\frac{e^{-M_{n'}\left(t_{\text{sep}}-\tau\right)}e^{-M_n\tau}}{4M_{n'}M_n}\times \\
  &\quad\quad\times Z_{n'}Z_n^\dagger\mathcal{P}^{\text{3pt}}_{\beta\alpha}u^{n'}_\alpha(\vec{0},s')\overline{u}^n_\beta(\vec{0},s)\times \\
  &\quad\quad\quad\quad\quad\quad\quad\times\bra{n',p',s'}\mathcal{J}\ket{n,p,s} \\
  &=\sum_{n,s}\sum_{n',s'}\frac{e^{-M_{n'}\left(t_{\text{sep}}-\tau\right)}e^{-M_n\tau}}{4M_{n'}M_n}\times \\
  &\quad\quad\times Z_{n'}Z_n^\dagger\mathcal{P}^{\text{3pt}}_{\beta\alpha}u^{n'}_\alpha(\vec{0},s')\overline{u}^n_\beta(\vec{0},s)\times \\
  &\quad\quad\quad\quad\quad\quad\quad\times\overline{u}^{n'}_\rho(\vec{0},s')\mathcal{J}_{\rho\sigma}u_\sigma(\vec{0},s) \\
  &=\sum_{n',n}\frac{Z_{n'}Z_n^\dagger}{4M_{n'}M_n}e^{-M_{n'}\left(t_{\text{sep}}-\tau\right)}e^{-E_n\tau}\times \\
  &\quad\quad\times\text{Tr}\left[\mathcal{P}^{\text{3pt}}\left(-i\slashed{p}_E\thinspace'+M_{n'}\right)\mathcal{J}_{n'n}\left(-i\slashed{p}_E+M_n\right)\right].
\end{align*}
It then follows $\mathcal{A}=2g_\Gamma^{00}\left|Z_0\right|^2$ and $\mathcal{B}=2g_\Gamma^{11}\left|Z_1\right|^2$. We can then extract the ground-state matrix element via
\begin{equation*}
  g_{00}^\Gamma=\mathcal{A}/\left|\mathbf{a}\right|^2.
\end{equation*}
To extract the masses, overlap factors and matrix elements from our
data, we perform simultaneous fits to the two-point and three-point
correlation functions according to Equations \ref{eq:2ptfit} and
\ref{eq:3ptfit}, respectively.

\section{Results}\label{sec:results}
\begin{figure}[h!]
  \centering
  \includegraphics[width=0.52\textwidth]{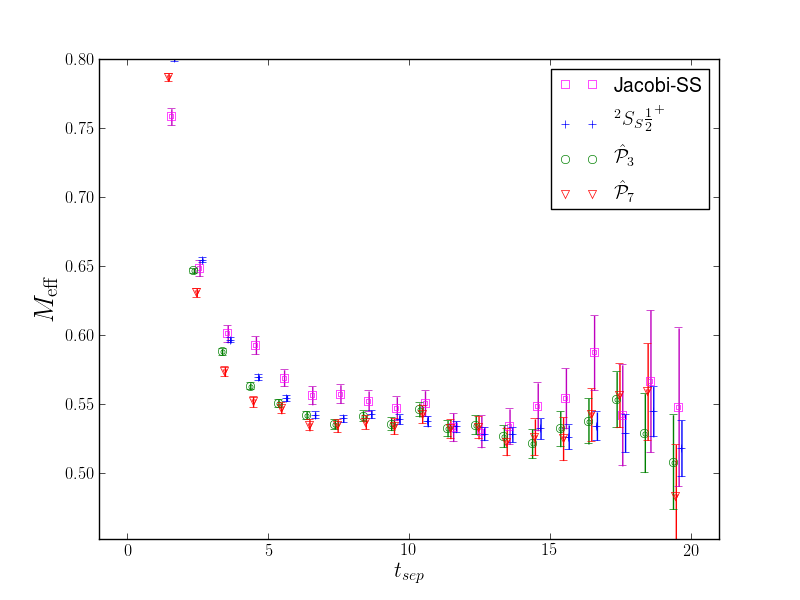}
  \caption{Nucleon effective mass when using Jacobi-SS (purple) and distilled interpolators. The non-relativistic $\thinspace^2S_S\frac{1}{2}^+$ (blue) distilled interpolator shows considerable improvement over the Jacobi-SS interpolator, while applying the GEVP to a basis of distilled interpolators of order 3 (green) and of order 7 (red) show further improvement.} 
  \label{fig:meff_t2-16}
\end{figure}

We compute the two-point functions averaged over three source
positions.  The extraction for the matrix elements is obtained from
three-point functions for $t_{\text{sep}}\in\lbrace8,12,16\rbrace$,
and $\tau\in\left[0,t_{\text{sep}}-1\right]$.
Rather than view the calculated two-point correlation functions directly, we judge the quality of the two-point correlators for each interpolator by plotting the effective mass
\begin{equation}
  M_{\mathrm{eff}}\left(t+0.5\right)=\frac{1}{a}\ln{\frac{C^{2\text{pt}}\left(t\right)}{C^{2\text{pt}}\left(t+1\right)}}
\end{equation}
as a function of the source-sink separation $t$. Calculating each two-point correlator, averaged over three different source positions, leads to the effective masses seen in Figure \ref{fig:meff_t2-16}. The lack of a plateau in the effective mass of the Jacobi-SS interpolator until $t\sim10$ is indicative of excited-state contamination for source-sink separations of $\lesssim1\text{ fm}$. Use of the $\thinspace^2S_S\frac{1}{2}^+$ distilled interpolator leads to an earlier onset of a plateau in the effective mass, with statistical uncertainty that is at least $50\%$ smaller than that of the Jacobi-SS interpolator. This plateau is seen to begin for $t\sim6$, or $\sim0.6\text{ fm}$. It is also worth noting that the expected exponentially increasing noise in the nucleon effective mass is substantially suppressed at larger source-sink separations, when compared to the Jacobi-SS interpolator.

Use of a variationally improved interpolator derived from bases of distilled interpolators, ($\hat{\mathcal{P}}_3$ or $\hat{\mathcal{P}}_7$), leads to a more rapid exponential decay of excited states at early Euclidean times. The effective mass induced by the $\hat{\mathcal{P}}_3$ interpolator exhibits a plateau that has nearly the same statistical precision as that of the $\thinspace^2S_S\frac{1}{2}^+$ interpolator, while the excited-states are seen to decay more rapidly for $t<5$. Expanding the basis of distilled interpolators, the $\hat{\mathcal{P}}_7$ interpolator leads to an even more rapid decay of excited states for $t<5$, consistent with a variationally improved interpolator receiving excited-state contributions from states higher in energy than those within the basis. As with the $\thinspace^2S_S\frac{1}{2}^+$ and $\hat{\mathcal{P}}_3$ interpolators, the effective mass of $\hat{\mathcal{P}}_7$ begins around $t=6$, however the plateau is noticeably lower than the former. In general, the statistical precision of all distilled interpolators appears to be comparable, except for large Euclidean times wherein the determination of the variationally improved interpolators becomes increasingly uncertain. We attribute this increased uncertainty to stem from the variationally improved interpolators being relatively unconstrained at large source-sink separations, where elements of the correlation matrix (Eq. \ref{eq:corrmatrix}) are themselves dominated by noise.

\subsection{Two-State Analysis}
To quantify the discussion above and to guide our future simultaneous
fits to the two- and three-point functions, we explore the efficacy of
one- and two-state fits have in describing the calculated effective
masses. All fits are restricted such that $2\le t_{\text{fit}}\le16$
thereby avoiding contact terms in the clover Wilson action and a
collective fluctuation of the effective masses for $t\gtrsim16$. The
results of these fits are collected in Tables \ref{tab:1pt-fits} and
\ref{tab:2pt-fits}.
\begin{table}[h!]
  \centering
  \scriptsize{
  \begin{tabular}{ | c | c | c | c | c |}
    \hline
    \multicolumn{5}{|c|}{Fit $C^{2\text{pt}}\left(t\right)=\left|\mathbf{a}\right|^2e^{-M_0t}$}\\
    \hline
    $\hat{\mathcal{O}}$ & $t_{\text{fit}}$ & $\left|\mathbf{a}\right|^2$ & $M_0$ & $\chi_r^2$ \\
    \hline
    \multirow{4}{*}{\tiny Jacobi-SS} & $\left[4,16\right]$ & 5.032(095)e-08 & 0.563(4) & 4.334 \\
    & $\left[5,16\right]$ & 4.701(100)e-08 & 0.556(3) & 1.458 \\
    & $\left[6,16\right]$ & 4.516(122)e-08 & 0.551(4) & 0.932 \\
    & $\left[7,16\right]$ & 4.450(148)e-08 & 0.550(4) & 0.977 \\ \hline
    \multirow{4}{*}{\tiny $\thinspace^2S_S\frac{1}{2}^+$} & $\left[4,16\right]$ & 1.627(011)e-02 & 0.548(1) & 20.77 \\
    & $\left[5,16\right]$ & 1.552(011)e-02 & 0.542(1) & 6.314 \\
    & $\left[6,16\right]$ & 1.500(013)e-02 & 0.538(1) & 2.354 \\
    & $\left[7,16\right]$ & 1.483(016)e-02 & 0.537(1) & 2.237 \\ \hline
    \multirow{4}{*}{\tiny $\hat{\mathcal{P}}_3$} & $\left[4,16\right]$ & 1.159(08)e+00 & 0.546(1) & 9.819 \\
    & $\left[5,16\right]$ & 1.114(09)e+00 & 0.541(1) & 3.227 \\
    & $\left[6,16\right]$ & 1.084(11)e+00 & 0.538(1) & 1.510 \\
    & $\left[7,16\right]$ & 1.073(13)e+00 & 0.537(2) & 1.393 \\ \hline
    \multirow{4}{*}{\tiny $\hat{\mathcal{P}}_7$} & $\left[4,16\right]$ & 1.045(09)e+00 & 0.540(1) & 3.094 \\
    & $\left[5,16\right]$ & 1.021(10)e+00 & 0.537(2) & 1.509 \\
    & $\left[6,16\right]$ & 0.999(12)e+00 & 0.535(2) & 0.613 \\
    & $\left[7,16\right]$ & 0.999(15)e+00 & 0.535(2) & 0.688 \\ \hline
  \end{tabular}
  }
  \caption{Parameters of a single-state fit to the two-point correlators. As
    the initial time slice over which fits are performed is made
    larger, the determined ground-state mass, as expected, is found to
    decrease. The ground-state mass of the Jacobi-SS interpolator is
    seen to converge toward $M_0\sim0.55$, while for the distilled
    interpolators the ground-state mass appears to converge towards
    $M_0\sim0.535$. Errors are purely statistical.}
  \label{tab:1pt-fits}
\end{table}

Although single-state fits make explicit the improvements gained by using distillation over standard Jacobi-smearing, notably a $\sim3\%$ difference in the determination of the ground-state mass, the improvements at this stage hardly seem worth the cost of constructing the required distillation basis. By including a second state in the fits performed to the two-point correlators, the gains produced by distillation are quite encouraging.

\begin{table*}[h!]
  \scriptsize{
  \begin{tabular}{ | c | c | c | c | c | c | c |}
    \hline
    \multicolumn{7}{|c|}{Fit $C^{2\text{pt}}\left(t\right)=\left|\mathbf{a}\right|^2e^{-M_0t}+\left|\mathbf{b}\right|^2e^{-M_1t}$}\\
    \hline
    $\hat{\mathcal{O}}$ & $t_{\text{fit}}$ & $\left|\mathbf{a}\right|^2$ & $M_0$ & $\left|\mathbf{b}\right|^2$ & $M_1$ & $\chi_r^2$ \\
    \hline
    \multirow{3}{*}{\tiny Jacobi-SS} & $\left[2,16\right]$ & 4.12(25)e-08 & 0.543(6) & 3.70(25)e-08 & 1.04(08) & 0.842 \\
    & $\left[3,16\right]$ & 3.81(42)e-08 & 0.536(9) & 3.12(35)e-08 & 0.91(11) & 0.753  \\
    & $\left[4,16\right]$ & 4.14(48)e-08 & 0.54(01) & 5.3(5.9)e-08 & 1.13(42) & 0.667 \\ \hline
    \multirow{3}{*}{\tiny $\thinspace^2S_S\frac{1}{2}^+$} & $\left[2,16\right]$ & 1.45(02)e-02 & 0.536(2) & 1.69(06)e-02 & 1.25(03) & 1.535 \\
    & $\left[3,16\right]$ & 1.43(03)e-02 & 0.534(2) & 1.35(14)e-02 & 1.14(06) & 1.268 \\
    & $\left[4,16\right]$ & 1.42(04)e-02 & 0.534(3) & 1.31(52)e-02 & 1.13(15) & 1.407 \\ \hline
    \multirow{3}{*}{\tiny $\hat{\mathcal{P}}_3$} & $\left[2,16\right]$ & 1.07(1)e+00 & 0.536(2) & 1.21(7)e+00 & 1.32(04) & 1.159 \\
    & $\left[3,16\right]$ & 1.05(2)e+00 & 0.535(2) & 0.935(157) & 1.21(08) & 1.069 \\
    & $\left[4,16\right]$ & 1.05(3)e+00 & 0.535(2) & 1.00(63)e+00 & 1.23(21) & 1.185 \\ \hline
    \multirow{3}{*}{\tiny $\hat{\mathcal{P}}_7$} & $\left[2,16\right]$ & 1.00(1)e+00 & 0.535(2) & 1.08(13)e+00 & 1.43(08) & 0.737 \\
    & $\left[3,16\right]$ & 0.98(2)e+00 & 0.533(2) & 0.68(24)e+00 & 1.23(17) & 0.668 \\
    & $\left[4,16\right]$ & 0.98(4)e+00 & 0.533(3) & 0.48(53)e+00 & 1.13(39) & 0.729 \\ \hline
  \end{tabular}
  }
  \caption{Parameters of two-state fits to the two-point correlators. There is general agreement in determination of the ground-state mass, while the mass gap $\Delta m=M_1-M_0$ is seen to become ever larger as distillation, and the GEVP applied to a basis of distilled interpolators, in place of the Jacobi-SS interpolator. Fits performed over larger source-sink separations were increasingly consistent with single-state fits, and are not considered further. Errors are purely statistical.}
  \label{tab:2pt-fits}
\end{table*}
The inclusion of a second state in fits to the two-point correlators yields slightly smaller determinations for the ground-state mass and overlap factors when using the Jacobi, $\thinspace^2S_S\frac{1}{2}^+$, and $\hat{\mathcal{P}}_3$ interpolators, while little change is seen in the $\hat{\mathcal{P}}_7$ interpolator. The most striking difference between the interpolators comes in the determination of the first-excited state mass. The determined mass gaps are 
\begin{equation*}
  \Delta m=\left(M_1-M_0\right)\simeq
  \begin{cases}
    0.497(80) & \quad \text{Jacobi-SS} \\
    0.714(30) & \quad \thinspace^2S_S\frac{1}{2}^+ \\
    0.784(40) & \quad \hat{\mathcal{P}}_3 \\
    0.895(80) & \quad \hat{\mathcal{P}}_7
  \end{cases}
\end{equation*}
Evidently distillation and the variational method lead to greater elimination of excited-state contributions to the two-point correlators, where the mass gap is $\sim44\%$, $\sim58\%$, and $\sim80\%$ larger for the $\thinspace^2S_S\frac{1}{2}^+$, $\hat{\mathcal{P}}_3$, and $\hat{\mathcal{P}}_7$ interpolators, respectively, when compared to the Jacobi-SS interpolator. The compounded improvements of the variational method applied to our bases of distilled interpolators is entirely consistent with \cite{Blossier:2009kd}.

\subsection{Results for $g_S^{u-d}$, $g_A^{u-d}$, $g_T^{u-d}$, $g_V^{u-d}$}
In this section we seek a more reliable determination of the masses, overlap factors, and likewise nucleon matrix elements, by performing simultaneous fits to three-point and two-point correlators with interpolator $\mathcal{O}$ and current structure $\Gamma$. We fix the window over which the two-point correlators are fit to be $t_{2\text{pt}}^{\text{fit}}\in\left[2,16\right]$, while several three-point fit windows are studied. The three-point fit windows are identified by $\tau_{\text{buff}}$ - that is for a given $t_{\text{sep}}$, $\tau_{\text{fit}}\in\left[0+\tau_{\text{buff}},t_{\text{sep}}-\tau_{\text{buff}}\right]$.

To illustrate the extracted isovector charges and to quantify the
degree of excited-state contamination, we plot an effective charge
defined as
\begin{equation}
  g_{\Gamma\text{, eff}}\left(t_{\text{sep}},\tau\right)=\frac{C_\Gamma^{\text{3pt}}\left(t_{\text{sep}},\tau\right)}{C_{\text{fit}}^{\text{2pt}}\left(t_{\text{sep}}\right)}
\end{equation}
where $C_\Gamma^{\text{3pt}}\left(t_{\text{sep}},\tau\right)$ is the
three-point correlation function for a given source-sink separation
and current insertion time, and
$C_{\text{fit}}^{\text{2pt}}\left(t_{\text{sep}}\right)$ is the
corresponding best fit applied to the two-point function of the same
interpolators and source-sink separation. Errors on the effective
charges are computed via a simultaneous jackknife re-sampling of the
two-point fit and three-point correlator, to account for the
correlations between fitted parameters. Shown together with the
effective charges are our extracted values for the isovector charges
using the methodology described above. Error (gray) bands of the
extracted charges (black line) are determined through the ratio of fit
parameters $\mathcal{A}/\left|\mathbf{a}\right|^2$, likewise
accounting for parameter covariance. We highlight that nearly the same
masses for the ground- and first-excited states are found from the
2-state fits to the two-point correlators and from the simultaneous
fits to the two- and three-point correlators.

\subsubsection{$g_S^{u-d}$}
Coupling of the isovector scalar current $S^3=\overline{q}\frac{\tau^3}{2}q$ to the Jacobi-SS interpolator $\mathcal{N}_{\alpha}$ produces an effective matrix element (denoted here as $\mathcal{M}_S^{\text{eff}}$) that is determined to within no less than $\sim10\%$ uncertainty. Although $\mathcal{M}_S^{\text{eff}}$ is symmetric about the midpoint $\tau-t_{\text{sep}}/2$ for $t_{\text{sep}}=8$, indicating equal excited-state contamination on the source/sink side of the insertion, $\mathcal{M}_S^{\text{eff}}$ exhibits antisymmetric behavior for $t_{\text{sep}}/a=12$ and is largely statistical noise for source-sink separations greater than $1.5\text{ fm}$.

Immediately noticeable with the use of a single distilled interpolator
($\thinspace^2S_S\frac{1}{2}^+$) is the considerable reduction in
statistical uncertainty of $\mathcal{M}_S^{\text{eff}}$ at each value
of $t_{\text{sep}}$. Moreover, the determinations of
$\mathcal{M}_S^{\text{eff}}$ for different $t_{\text{sep}}$ are in
greater agreement, with a clear plateau established for
$t_{\text{sep}}\sim1.18\text{ fm}$. Recalling that a simultaneous fit
for all values of $t_{\rm sep}$ has been performed to extract the
matrix element, the fact that the $t_{\text{sep}}=16$ values of
$\mathcal{M}_S^{\text{eff}}$ are not necessarily on the curve reflects
that the fit is largely constrained by data with $t_{\text{sep}}\lesssim12$.

Introduction of two hybrid interpolators to construct $\hat{\mathcal{P}}_3$ appears to return the $t_{\text{sep}}=8,12$ determinations of $\mathcal{M}_S^{\text{eff}}$ to a form structurally similar to that found for the Jacobi-SS interpolator. It is curious to note that the $t_{\text{sep}}=16$ determination of $\mathcal{M}_S^{\text{eff}}$ is nearly identical to the corresponding determination with $\thinspace^2S_S\frac{1}{2}^+$, with inflated errors bars. These data suggest the inclusion of only the two hybrid interpolators provides limited improvement in the extraction of $g_S^{u-d}$. Inclusion of the remaining distilled interpolators in the variational analysis yields determinations of $\mathcal{M}_S^{\text{eff}}$ that not only exhibit broad plateaus as a function of $\tau$, but are also consistent within error and are consistent with the $\thinspace^2S_S\frac{1}{2}^+$ determination.

\subsubsection{$g_A^{u-d}$}
The isovector axial current
$A_\mu^3=\overline{q}\gamma_\mu\gamma_5\frac{\tau^3}{2}q$ is the
insertion that is most sensitive to excited-state contamination and
finite volume effects. As with $g_S^{u-d}$, determinations of the
effective matrix element (denoted here as
$\mathcal{M}_A^{\text{eff}}$) for different $t_{\text{sep}}$ using the
Jacobi-SS interpolator are plagued with poor statistics. Clearly
spatial Gaussian smearing of $\mathcal{N}_\alpha$ is not sufficient to
suppress excited-state effects.

Employing the local distilled interpolator $\thinspace^2S_S\frac{1}{2}^+$ yet again leads to dramatic reduction in statistical uncertainties in determinations $\mathcal{M}_A^{\text{eff}}$, and an increase in the extracted value of $g_A^{u-d}$ by $\sim7\%$. Notably broad plateaus in $\mathcal{M}_A^{\text{eff}}$ can likewise be seen for $t_{\text{sep}}=8,12$ that are consistent within error, thereby reducing the minimal source-sink separation ($\lesssim1\text{ fm}$) to reliably extract $g_A^{u-d}$.

The behavior of the $\hat{\mathcal{P}}_3$ interpolator is in line with
that found for $g_S^{u-d}$ - reduced agreement between
$t_{\text{sep}}=8,12$ determinations, and fits most heavily
constrained by data at smaller source-sink separations. Curiously, the
slight oscillation seen in the $t_{\text{sep}}=16$
$\thinspace^2S_S\frac{1}{2}^+$ determination of
$\mathcal{M}_A^{\mathrm{eff}}$ is amplified following application of
the variational method. Prior to $\tau-t_{\text{sep}}<-2$, the
$t_{\text{sep}}=16$ effective matrix element seems to fall into
agreement with the $t_{\text{sep}}=12$ determination. We speculate the
abrupt decrease in the $t_{\text{sep}}=16$ effective matrix element to
be caused by neglect of backward-propagating positive- parity states
in the corresponding two-point functions.

The $\hat{\mathcal{P}}_7$ interpolator perpetuates the double-plateau feature in the $t_{\text{sep}}=16$ determination of $\mathcal{M}_A^{\mathrm{eff}}$. Remarkably, however, application of the variational method to this enlarged basis of distilled interpolators demonstrates absolute agreement between the $t_{\text{sep}}=8,12$ determinations of $\mathcal{M}_A^{\mathrm{eff}}$.

\subsubsection{$g_T^{u-d}$}
The isovector tensor current $\mathcal{T}_{\mu\nu}^3=\overline{q}\frac{i}{2}\left[\gamma_\mu,\gamma_\nu\right]\frac{\tau^3}{2}q$ is among one of the best determined quantities in the nucleon, particularly at zero virtuality defining $g_T^{u-d}$. This facet is supported by models that show excitations of excited-states in the nucleon are suppressed when coupling the tensor current $\mathcal{T}^3_{\mu\nu}$ \cite{Roberts:2018}.

As with previous charges, the $\thinspace^2S_S\frac{1}{2}^+$ interpolator leads to a dramatic reduction in statistical uncertainty of the effective matrix elements (denoted here as $\mathcal{M}_T^{\mathrm{eff}}$) when compared to Jacobi-SS. By $t_{\text{sep}}\sim1.18\text{ fm}$, a definite plateau is present in $\mathcal{M}_T^{\mathrm{eff}}$ for several insertion times $\tau$; this same plateau is shared with the $t_{\text{sep}}=16$ determination. The absolute agreement of the central values and error of the $t_{\text{sep}}=12,16$ determinations suggest that when using distillation, a source-sink separation of roughly 1fm is sufficient to reliably extract $g_T^{u-d}$.

The variationally improved $\hat{\mathcal{P}}_3$ expands upon the improvements seen with the $\thinspace^2S_S\frac{1}{2}^+$ interpolator. Despite the extracted value of $g_T^{u-d}$ being roughly equal (1.147(13) vs. 1.145(15), respectively) for $\thinspace^2S_S\frac{1}{2}^+$ and $\hat{\mathcal{P}}_3$, the $t_{\text{sep}}=12,16$ determinations depict an even broader plateau that are again entirely consistent within error. These enlarged plateaus for numerous insertion times $\tau$ follow naturally from a determination of $\Delta m$ that is greater than that determined with $\thinspace^2S_S\frac{1}{2}^+$.

Proceeding to the $\hat{\mathcal{P}}_7$ interpolator continues the trends seen with other distilled interpolators, wherein the statistical uncertainties are greatly reduced and the plateaus in the effective matrix elements are seen to become ever larger. Remarkably, the $t_{\text{sep}}=16$ determination using $\hat{\mathcal{P}}_7$ resembles that of the vector charge - a matrix element that is constant in $\tau$ within minor statistical fluctuations. We emphasize here that the effective matrix element for $\tau-t_{\text{sep}}/2=-8$ is coincident with the source interpolator and thus should not be considered a reflection of $g_T^{u-d}$.

Recalling the functional form of the fit applied to the three-point correlator, Eq. \ref{eq:3ptfit}, the coefficient $\mathcal{B}$ encodes the first excited-state matrix element and $\mathcal{C}$ captures the ground to first excited-state transition matrix element. As evident from Table \ref{tab:tensorparams}, the determinations of $\mathcal{B}$ for each interpolator are generally consistent with zero, indeed supporting the notion that excited-state contributions to $g_T^{u-d}$ are greatly suppressed. On the other hand, the largest source of contamination appears to stem from the transition matrix element contained in $\mathcal{C}$.

\subsubsection{$g_V^{u-d}$}
As the isovector vector current $V^3_\mu=\overline{q}\gamma_\mu\frac{\tau^3}{2}q$ simply counts the number of quarks within the nucleon, and all its excitations, there is no surprise in the extraction of $g_V^{u-d}$ for each interpolator. We have included the nucleon vector charge here as a useful sanity check, thus ensuring such an observable that is independent of state and $\tau$ is indeed recovered. The continuum requirement $1=Z_Vg_V^{u-d}$ trivially sets the vector current renormalization to be $Z_V\sim0.856$, consistent with determinations on finer lattice ensembles with lighter $m_\pi$ \cite{Yoon:2016dij,Yoon:2016jzj}. As for the previous charges, distillation alone affords considerably improved statistics over Jacobi-SS.  That said, application of the variational method to a basis of distilled interpolators appears to produce a curious ``spreading'' in the determined effective vector matrix element.

\begin{figure*}[tb]
    \centering
    \includegraphics[width=0.49\linewidth]{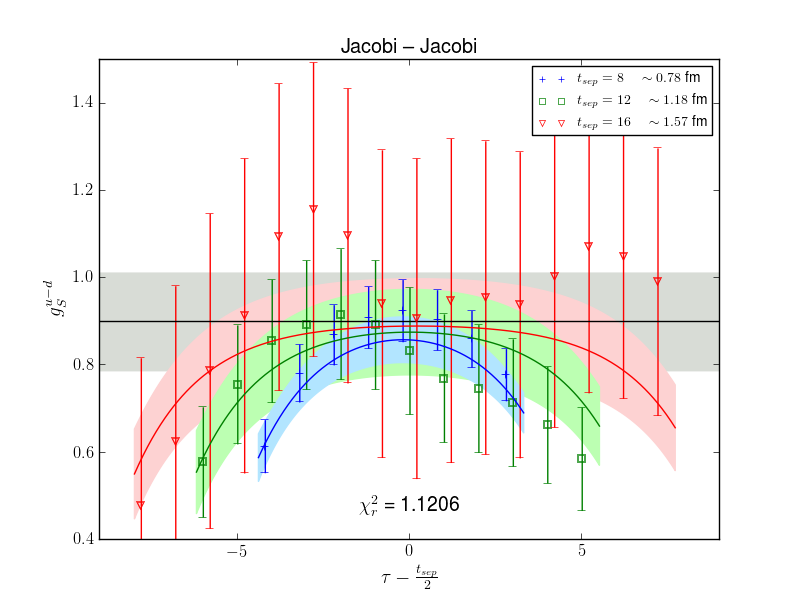}
    \centering
    \includegraphics[width=0.49\linewidth]{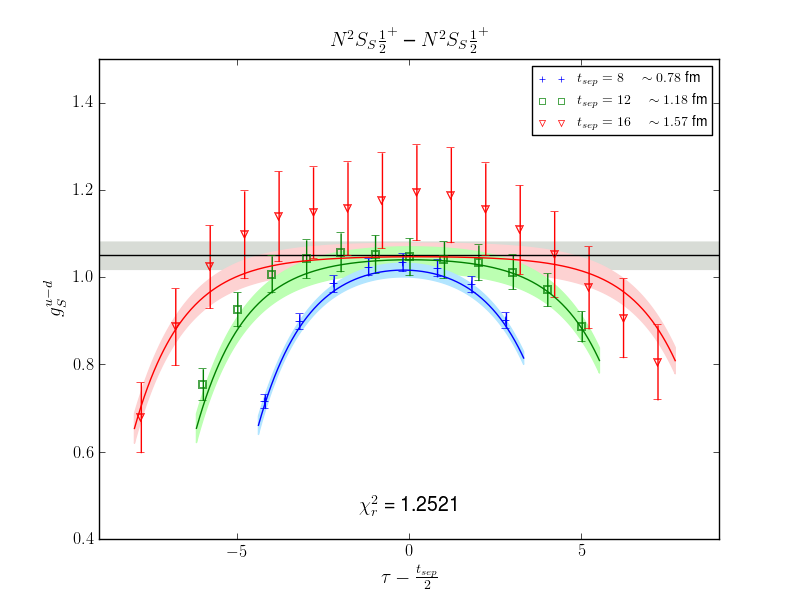}
  \caption{Extracted effective unrenormalized isovector scalar charge using the Jacobi-SS (left) and the $\thinspace^2S_S\frac{1}{2}^+$ (right) distilled interpolators. Displayed plots are for simultaneous fits with $t^{\text{fit}}_{2\text{pt}}\in\left[2,16\right]$ and $\tau_{\text{buff}}=2$.}
\end{figure*}
\begin{figure*}[tb]
    \centering
    \includegraphics[width=0.49\linewidth]{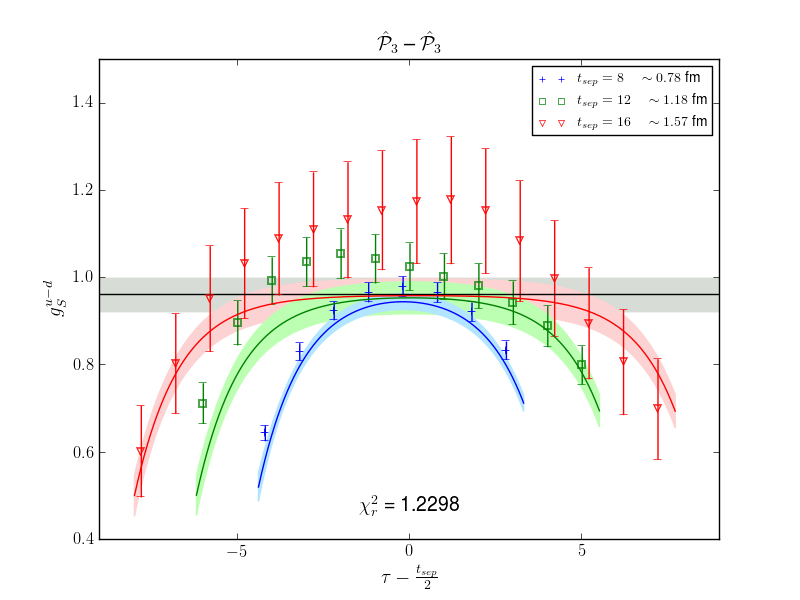}
    \includegraphics[width=0.49\linewidth]{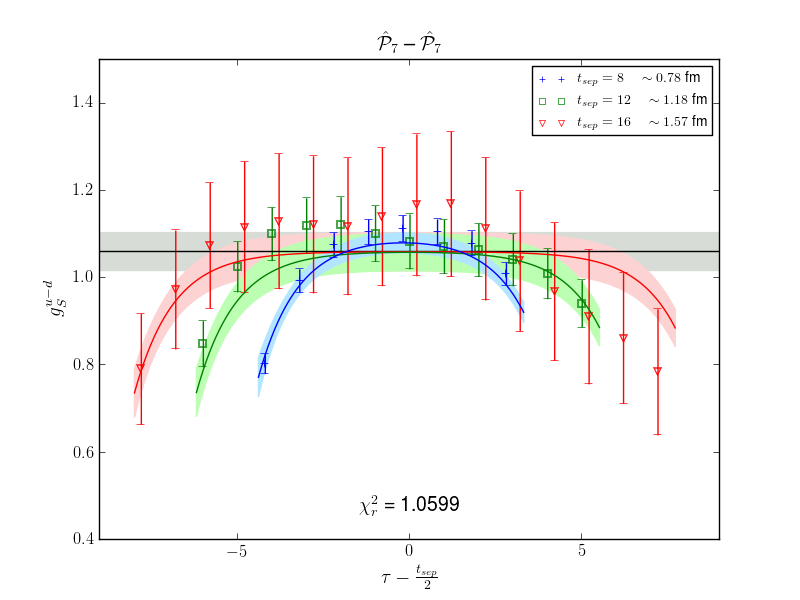}
  \caption{Extracted effective unrenormalized isovector scalar charge using the projected distilled interpolator from the 3x3 GEVP (left) and the projected distilled interpolator from the 7x7 GEVP (right). Displayed plots are for simultaneous fits with $t^{\text{fit}}_{2\text{pt}}\in\left[2,16\right]$ and $\tau_{\text{buff}}=2$.}
\end{figure*}
\begin{table*}[tb]
  \centering
  \scriptsize{
    \begin{tabular}{ | c | c | c | c | c | c | c | c | c | c |}
      \hline
      $\hat{\mathcal{O}}$ & $\tau_{\text{buff}}$ & $\mathcal{A}$ & $\mathcal{B}$ & $\mathcal{C}$ & $M_0$ & $M_1$ & $\left|\mathbf{a}\right|^2$ & $\left|\mathbf{b}\right|^2$ & $g^{u-d}_{S,\text{bare}}$ \\
      \hline
      \multirow{4}{*}{Jacobi-SS} & 1 & 3.72(40)e-08 & 2.57(3.12)e-07 & -2.93(28)e-08 & 0.546(5) & 1.087(77) & 4.26(20)e-08 & 3.72(26)e-08 & 0.87(10) \\
      & 2 & 3.75(44)e-08 & 1.33(2.46)e-07 & -2.63(47)e-08 & 0.544(5) & 1.061(76) & 4.17(22)e-08 & 3.71(25)e-08 & 0.90(11) \\
      & 3 & 3.86(51)e-08 & 1.54(2.87)e-07 & -3.08(95)e-08 & 0.543(6) & 1.054(78) & 4.13(23)e-08 & 3.75(25)e-08 & 0.94(13) \\
      & 4 & 3.28(65)e-08 & 2.08(3.09)e-07 & 0.34(3.47)e-08 & 0.543(6) & 1.044(75) & 4.11(23)e-08 & 3.75(24)e-08 & 0.80(16) \\ \hline
      \multirow{4}{*}{$\thinspace^2S_S\frac{1}{2}^+$} & 1 & 1.54(04)e-02 & 0.48(1.08)e-01 & -9.97(27)e-03 & 0.537(1) & 1.246(26) & 1.46(02)e-02 & 1.67(05)e-02 & 1.051(28) \\
      & 2 & 1.54(05)e-02 & 0.47(1.23)e-01 & -1.00(06)e-02 & 0.537(1) & 1.249(28) & 1.46(02)e-02 & 1.68(05)e-02 & 1.051(31) \\
      & 3 & 1.52(05)e-02 & 1.27(1.38)e-01 & -1.05(13)e-02 & 0.537(1) & 1.251(28) & 1.46(02)e-02 & 1.69(06)e-02 & 1.042(34) \\
      & 4 & 1.52(06)e-02 & 0.18(1.41)e-01 & -4.3(5.5)e-03 & 0.536(1) & 1.247(28) & 1.46(02)e-02 & 1.68(06)e-02 & 1.043(40) \\ \hline
      \multirow{4}{*}{$\hat{\mathcal{P}}_3$} & 1 & 1.01(4)e+00 & 9.8(14.0)e+00 & -7.44(23)e-01 & 0.537(1) & 1.289(35) & 1.06(1)e+00 & 1.15(5)e+00 & 0.952(35) \\
      & 2 & 1.03(4)e+00 & 11.7(19.4)e+00 & -8.43(64)e-01 & 0.538(1) & 1.317(40) & 1.07(1)e+00 & 1.19(6)e+00 & 0.961(38) \\
      & 3 & 1.03(5)e+00 & 27.7(25.8)e+00 & -1.06(15)e+00 & 0.538(1) & 1.328(41) & 1.07(1)e+00 & 1.21(6)e+00 & 0.958(43) \\
      & 4 & 1.05(5)e+00 & 8.2(24.8)e+00 & -0.77(64)e+00 & 0.537(2) & 1.324(41) & 1.07(1)e+00 & 1.21(6)e+00 & 0.984(49) \\ \hline
      \multirow{4}{*}{$\hat{\mathcal{P}}_7$} & 1 & 1.08(5)e+00 & 30.5(65.3)e+00 & -6.09(29)e-01 & 0.536(2) & 1.440(69) & 1.00(1)e+00 & 1.09(11)e+00 & 1.079(48) \\
      & 2 & 1.06(4)e+00 & 42.4(48.4)e+00 & -5.43(80)e-01 & 0.536(2) & 1.416(71) & 1.00(1)e+00 & 1.06(11)e+00 & 1.061(43) \\
      & 3 & 1.05(5)e+00 & 94.8(73.0)e+00 & -0.73(20)e+00 & 0.536(2) & 1.442(72) & 1.00(1)e+00 & 1.10(12)e+00 & 1.051(46) \\
      & 4 & 1.05(6)e+00 & 57.0(78.0)e+00 & 0.6(1.1)e+00 & 0.536(2) & 1.443(77) & 1.00(1)e+00 & 1.10(12)e+00 & 1.045(55) \\ \hline
    \end{tabular}
  }
  \caption{Results of simultaneous fits to the two-point and three-point correlators with scalar current insertion. The functional forms are given in Eqns. \ref{eq:2ptfit} and \ref{eq:3ptfit}.}
\end{table*}

\begin{figure*}[tb]
  \centering
    \centering
    \includegraphics[width=0.49\linewidth]{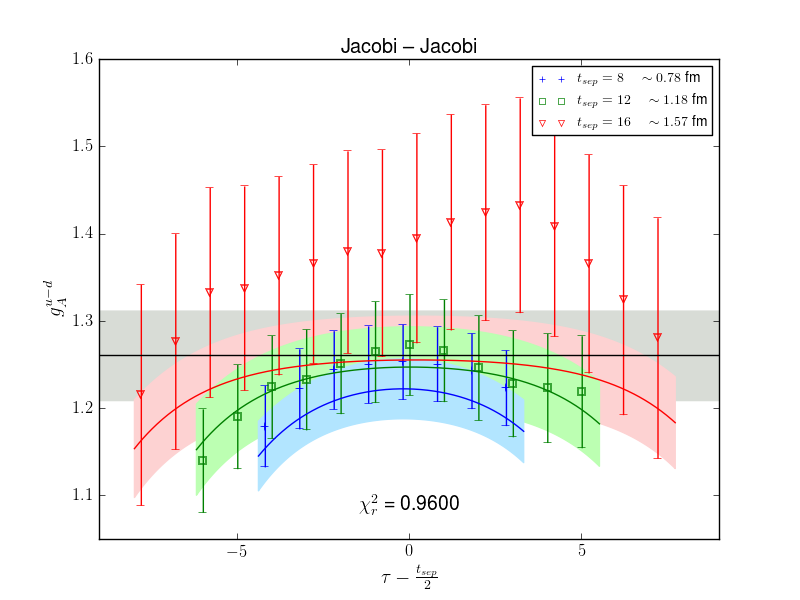}
    \includegraphics[width=0.49\linewidth]{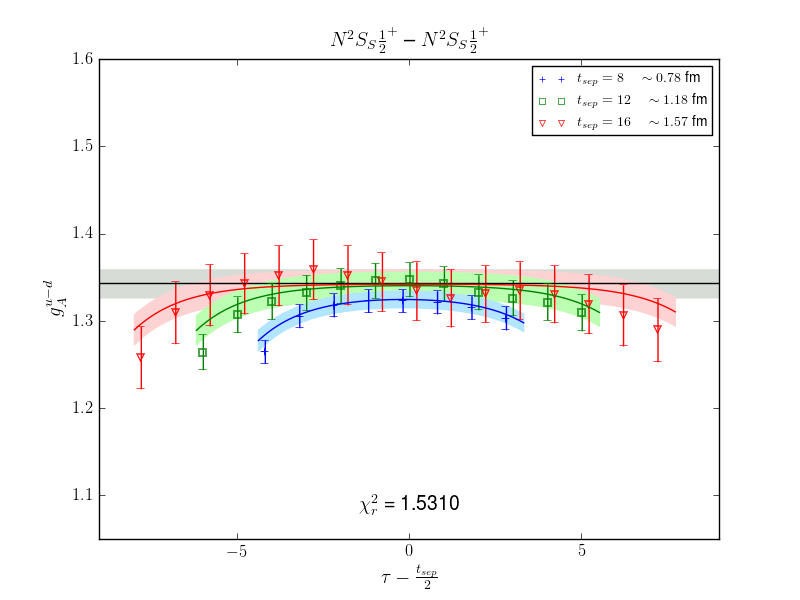}
  \caption{Extracted effective unrenormalized isovector axial charge using the Jacobi-SS (left) and the $\thinspace^2S_S\frac{1}{2}^+$ (right) distilled interpolators. Displayed plots are for simultaneous fits with $t^{\text{fit}}_{2\text{pt}}\in\left[2,16\right]$ and $\tau_{\text{buff}}=2$.}
\end{figure*}
\begin{figure*}[tb]
  \centering
    \centering
    \includegraphics[width=0.49\linewidth]{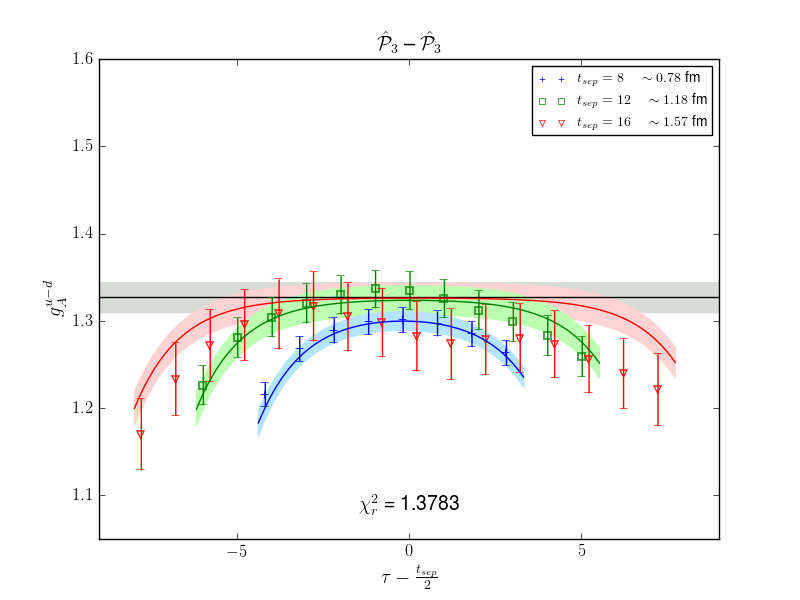}
    \includegraphics[width=0.49\linewidth]{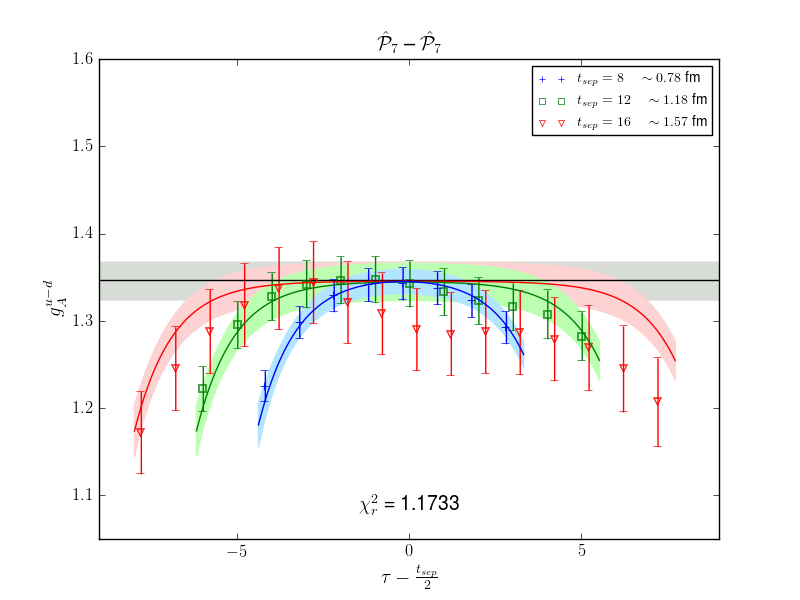}
  \caption{Extracted effective unrenormalized isovector axial charge using the projected distilled interpolator from the 3x3 GEVP (left) and the projected distilled interpolator from the 7x7 GEVP (right). Displayed plots are for simultaneous fits with $t^{\text{fit}}_{2\text{pt}}\in\left[2,16\right]$ and $\tau_{\text{buff}}=2$.}
\end{figure*}
\begin{table*}[tb]
  \centering
  \scriptsize{
    \begin{tabular}{ | c | c | c | c | c | c | c | c | c | c |}
      \hline
      $\hat{\mathcal{O}}$ & $\tau_{\text{buff}}$ & $\mathcal{A}$ & $\mathcal{B}$ & $\mathcal{C}$ & $M_0$ & $M_1$ & $\left|\mathbf{a}\right|^2$ & $\left|\mathbf{b}\right|^2$ & $g^{u-d}_{A,\text{bare}}$ \\
      \hline
      \multirow{4}{*}{Jacobi-SS} & 1 & 4.92(32)e-08 & 2.65(6.23)e-08 & -6.66(1.32)e-09 & 0.539(5) & 0.995(62) & 3.93(23)e-08 & 3.76(20)e-08 & 1.253(52) \\
      & 2 & 5.00(33)e-08 & 3.23(7.15)e-08 & -7.66(2.12)e-09 & 0.540(6) & 1.010(67) & 3.96(23)e-08 & 3.81(21)e-08 & 1.261(51) \\
      & 3 & 5.03(33)e-08 & 6.83(6.78)e-08 & -1.30(0.44)e-08 & 0.540(6) & 0.999(66) & 3.95(23)e-08 & 3.74(21)e-08 & 1.272(54) \\
      & 4 & 5.37(36)e-08 & 1.54(1.15)e-07 & -4.00(1.69)e-08 & 0.542(6) & 1.032(75) & 4.08(24)e-08 & 3.70(23)e-08 & 1.315(59) \\ \hline
      \multirow{4}{*}{$\thinspace^2S_S\frac{1}{2}^+$} & 1 & 1.95(03)e-02 & -2.52(5.57)e-02 & -1.38(0.11)e-03 & 0.536(1) & 1.242(28) & 1.46(02)e-02 & 1.67(05)e-02 & 1.342(16) \\
      & 2 & 1.96(03)e-02 & -3.03(5.78)e-02 & -1.33(0.23)e-03 & 0.536(1) & 1.244(28) & 1.46(02)e-02 & 1.68(05)e-02 & 1.343(16) \\
      & 3 & 1.96(03)e-02 & -1.77(5.85)e-02 & -1.70(0.54)e-03 & 0.536(1) & 1.247(28) & 1.46(02)e-02 & 1.68(06)e-02 & 1.344(16) \\
      & 4 & 1.98(04)e-02 & -0.55(6.32)e-02 & -6.11(2.61)e-03 & 0.536(1) & 1.253(28) & 1.46(02)e-02 & 1.69(06)e-02 & 1.357(17) \\ \hline
      \multirow{4}{*}{$\hat{\mathcal{P}}_3$} & 1 & 1.39(2)e+00 & -3.3(5.9)e+00 & -1.74(0.11)e-01 & 0.535(1) & 1.279(35) & 1.05(1)e+00 & 1.15(5)e+00 & 1.315(17) \\
      & 2 & 1.41(3)e+00 & -6.5(7.5)e+00 & -2.31(0.24)e-01 & 0.536(1) & 1.301(38) & 1.06(1)e+00 & 1.18(6)e+00 & 1.328(17) \\
      & 3 & 1.42(3)e+00 & -6.8(8.1)e+00 & -3.38(0.59)e-01 & 0.536(1) & 1.308(39) & 1.06(1)e+00 & 1.19(6)e+00 & 1.335(18) \\
      & 4 & 1.44(3)e+00 & -3.3(9.5)e+00 & -9.36(3.15)e-01 & 0.537(1) & 1.328(41) & 1.07(1)e+00 & 1.21(6)e+00 & 1.348(19) \\ \hline
      \multirow{4}{*}{$\hat{\mathcal{P}}_7$} & 1 & 1.33(3)e+00 & 12.5(18.2)e+00 & -2.51(0.13)e-01 & 0.535(1) & 1.403(58) & 9.96(11)e-01 & 1.04(9)e+00 & 1.339(20) \\
      & 2 & 1.35(3)e+00 & 11.5(26.9)e+00 & -2.87(0.38)e-01 & 0.535(2) & 1.442(72) & 9.99(12)e-01 & 1.10(12)e+00 & 1.347(22) \\
      & 3 & 1.35(3)e+00 & 16.4(27.8)e+00 & -2.94(0.91)e-01 & 0.535(2) & 1.451(76) & 1.00(1)e+00 & 1.11(13)e+00 & 1.345(21) \\
      & 4 & 1.35(3)e+00 & 23.7(29.5)e+00 & -6.39(5.15)e-01 & 0.535(2) & 1.443(77) & 1.00(1)e+00 & 1.10(12)e+00 & 1.346(22) \\ \hline
    \end{tabular}
  }
  \caption{Results of simultaneous fits to the two-point and three-point correlators with axial current insertion. The functional forms are given in Eqns. \ref{eq:2ptfit} and \ref{eq:3ptfit}.}
\end{table*}

\begin{figure*}[tb]
  \centering
    \centering
    \includegraphics[width=0.49\linewidth]{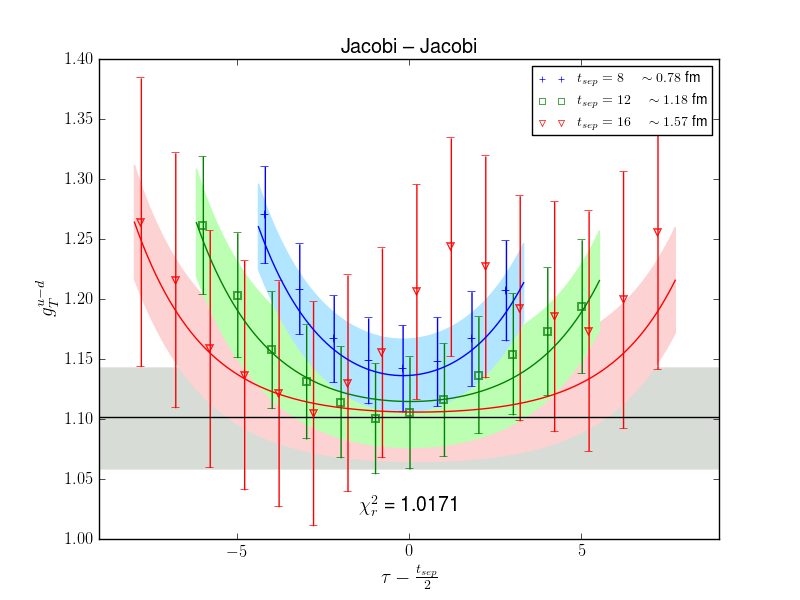}
    \includegraphics[width=0.49\linewidth]{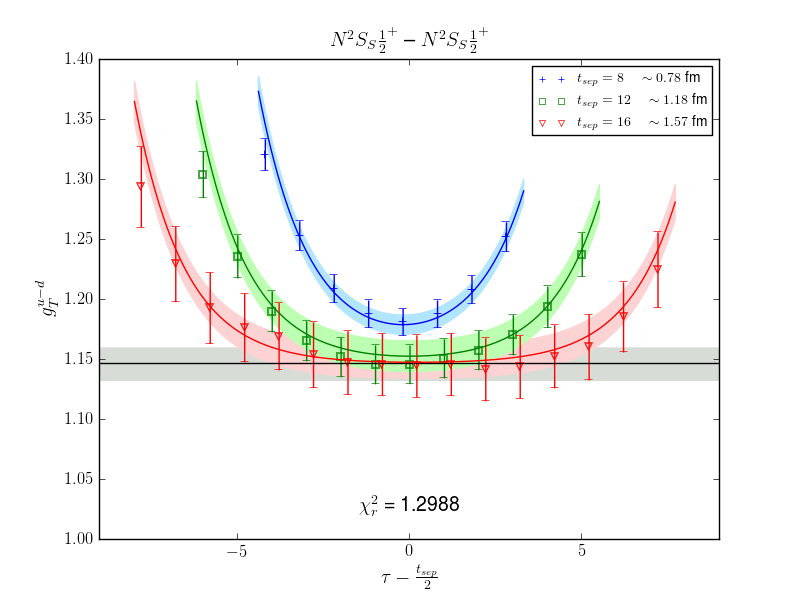}
  \caption{Extracted effective unrenormalized isovector tensor charge using the Jacobi-SS (left) and the $\thinspace^2S_S\frac{1}{2}^+$ (right) distilled interpolators. Displayed plots are for simultaneous fits with $t^{\text{fit}}_{2\text{pt}}\in\left[2,16\right]$ and $\tau_{\text{buff}}=2$.}
\end{figure*}
\begin{figure*}[tb]
  \centering
    \centering
    \includegraphics[width=0.49\linewidth]{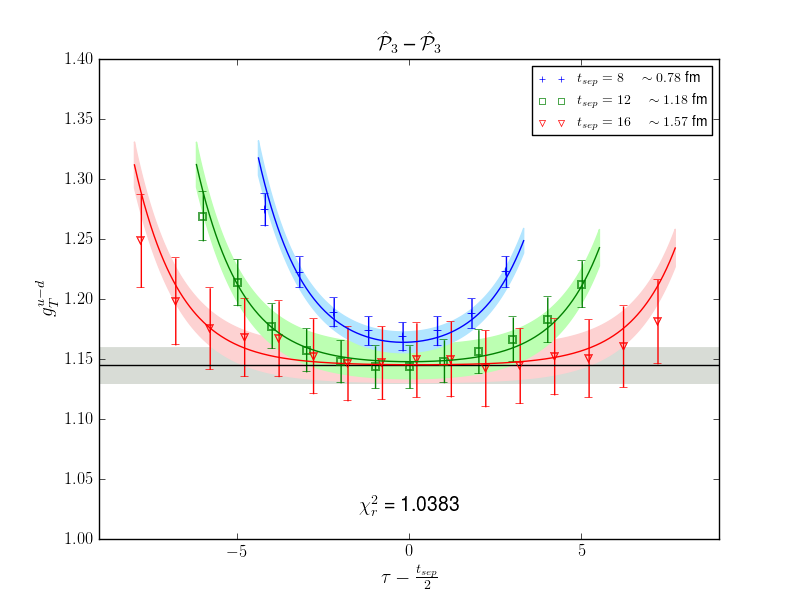}
    \includegraphics[width=0.49\linewidth]{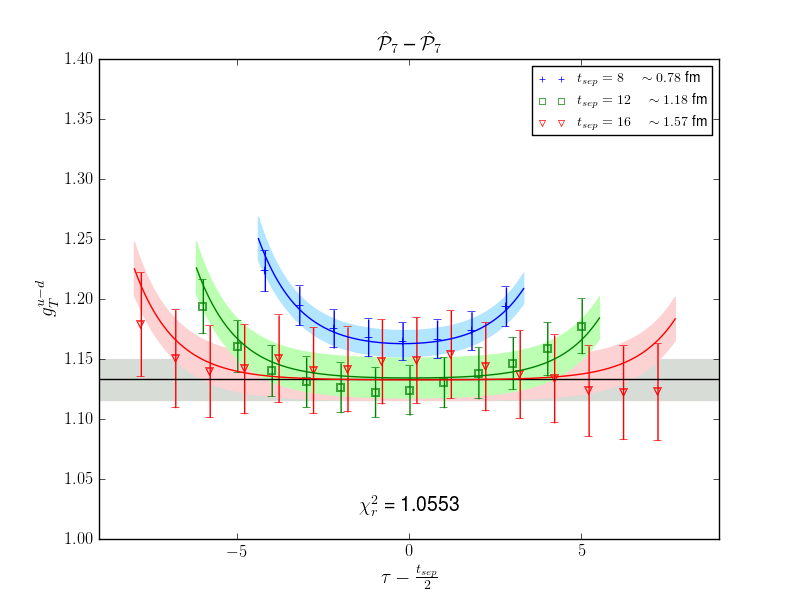}
  \caption{Extracted effective unrenormalized isovector tensor charge using the projected distilled interpolator from the 3x3 GEVP (left) and the projected distilled interpolator from the 7x7 GEVP (right). Displayed plots are for simultaneous fits with $t^{\text{fit}}_{2\text{pt}}\in\left[2,16\right]$ and $\tau_{\text{buff}}=2$.}
\end{figure*}
\begin{table*}[tb]
  \centering
  \scriptsize{
    \begin{tabular}{ | c | c | c | c | c | c | c | c | c | c |}
      \hline
      $\hat{\mathcal{O}}$ & $\tau_{\text{buff}}$ & $\mathcal{A}$ & $\mathcal{B}$ & $\mathcal{C}$ & $M_0$ & $M_1$ & $\left|\mathbf{a}\right|^2$ & $\left|\mathbf{b}\right|^2$ & $g^{u-d}_{T,\text{bare}}$ \\
      \hline
      \multirow{4}{*}{Jacobi-SS} & 1 & 4.70(29)e-08 & 0.21(1.28)e-07 & 1.39(0.11)e-08 & 0.547(5) & 1.113(71) & 4.25(19)e-08 & 3.95(28)e-08 & 1.105(41) \\
      & 2 & 4.49(32)e-08 & 3.36(7.57)e-08 & 1.21(0.19)e-08 & 0.542(6) & 1.043(73) & 4.07(23)e-08 & 3.78(23)e-08 & 1.101(42) \\
      & 3 & 4.48(33)e-08 & 5.02(7.93)e-08 & 1.31(0.39)e-08 & 0.543(6) & 1.045(74) & 4.10(23)e-08 & 3.77(24)e-08 & 1.093(43) \\
      & 4 & 4.50(33)e-08 & 2.45(9.79)e-08 & 2.04(1.28)e-08 & 0.544(6) & 1.059(79) & 4.15(23)e-08 & 3.76(26)e-08 & 1.084(45) \\ \hline
      \multirow{4}{*}{$\thinspace^2S_S\frac{1}{2}^+$} & 1 & 1.65(03)e-02 & 4.10(3.26)e-02 & 5.21(0.09)e-03 & 0.534(1) & 1.204(19) & 1.44(01)e-02 & 1.61(04)e-02 & 1.146(13) \\
      & 2 & 1.66(03)e-02 & 5.08(3.88)e-02 & 5.51(0.22)e-03 & 0.535(1) & 1.225(26) & 1.45(02)e-02 & 1.64(05)e-02 & 1.147(13) \\
      & 3 & 1.67(03)e-02 & 6.03(4.64)e-02 & 6.51(0.53)e-03 & 0.536(1) & 1.245(28) & 1.46(02)e-02 & 1.68(06)e-02 & 1.147(14) \\
      & 4 & 1.67(03)e-02 & 3.83(4.94)e-02 & 8.44(1.91)e-03 & 0.536(1) & 1.246(28) & 1.46(02)e-02 & 1.68(06)e-02 & 1.144(14) \\ \hline
      \multirow{4}{*}{$\hat{\mathcal{P}}_3$} & 1 & 1.22(2)e+00 & 4.31(5.36)e+00 & 2.94(0.08)e-01 & 0.536(1) & 1.292(31) & 1.06(1)e+00 & 1.16(4)e+00 & 1.144(15) \\
      & 2 & 1.22(2)e+00 & 4.19(5.74)e+00 & 3.05(0.20)e-01 & 0.536(1) & 1.301(38) & 1.06(1)e+00 & 1.17(5)e+00 & 1.145(15) \\
      & 3 & 1.22(2)e+00 & 4.53(6.51)e+00 & 3.79(0.48)e-01 & 0.537(2) & 1.316(40) & 1.07(1)e+00 & 1.20(6)e+00 & 1.145(15) \\
      & 4 & 1.22(3)e+00 & 1.87(7.17)e+00 & 6.62(2.24)e-01 & 0.537(2) & 1.319(41) & 1.07(1)e+00 & 1.20(6)e+00 & 1.139(16) \\ \hline
      \multirow{4}{*}{$\hat{\mathcal{P}}_7$} & 1 & 1.12(2)e+00 & 25.4(15.6)e+00 & 1.62(0.08)e-01 & 0.534(2) & 1.395(57) & 9.92(12)e-01 & 1.03(8)e+00 & 1.128(16) \\
      & 2 & 1.13(3)e+00 & 24.2(17.6)e+00 & 1.55(0.23)e-01 & 0.534(2) & 1.387(70) & 9.94(13)e-01 & 1.01(10)e+00 & 1.133(17) \\
      & 3 & 1.13(3)e+00 & 32.3(23.9)e+00 & 2.54(0.63)e-01 & 0.535(2) & 1.427(75) & 9.99(13)e-01 & 1.08(12)e+00 & 1.135(17) \\
      & 4 & 1.13(3)e+00 & 21.3(21.0)e+00 & 8.60(3.71)e-01 & 0.535(2) & 1.427(75) & 9.98(13)e-01 & 1.08(12)e+00 & 1.127(18) \\ \hline
    \end{tabular}
  }
  \caption{Results of simultaneous fits to the two-point and three-point correlators with tensor current insertion. The functional forms are given in Eqns. \ref{eq:2ptfit} and \ref{eq:3ptfit}.}
  \label{tab:tensorparams}
\end{table*}

\begin{figure*}[tb]
  \centering
    \includegraphics[width=0.49\linewidth]{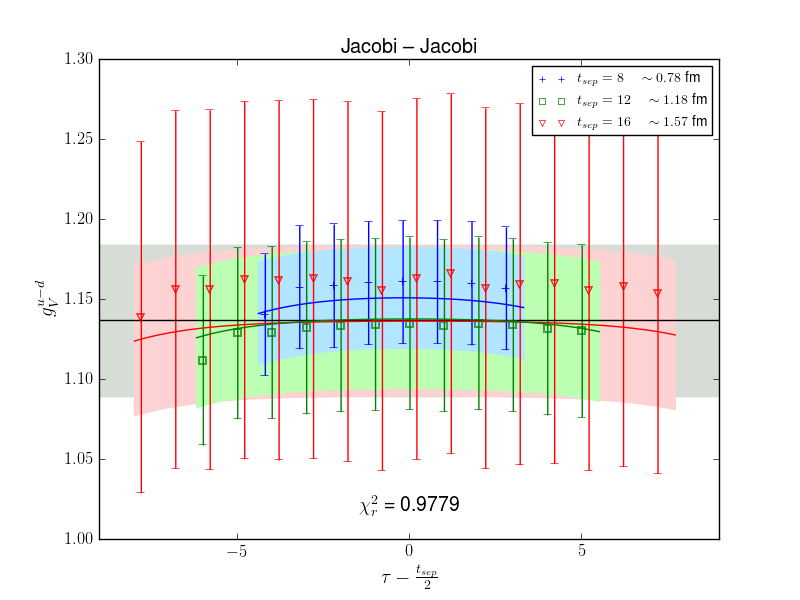}
    \includegraphics[width=0.49\linewidth]{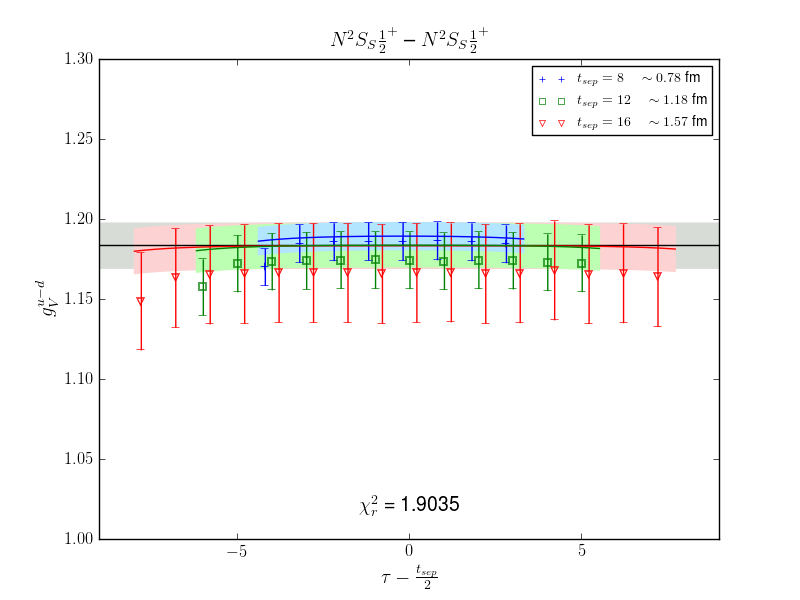}
  \caption{Extracted effective unrenormalized isovector vector charge using the Jacobi-SS (left) and the $\thinspace^2S_S\frac{1}{2}^+$ (right) distilled interpolators. Displayed plots are for simultaneous fits with $t^{\text{fit}}_{2\text{pt}}\in\left[2,16\right]$ and $\tau_{\text{buff}}=2$.}
\end{figure*}
\begin{figure*}[tb]
  \centering
    \includegraphics[width=0.49\linewidth]{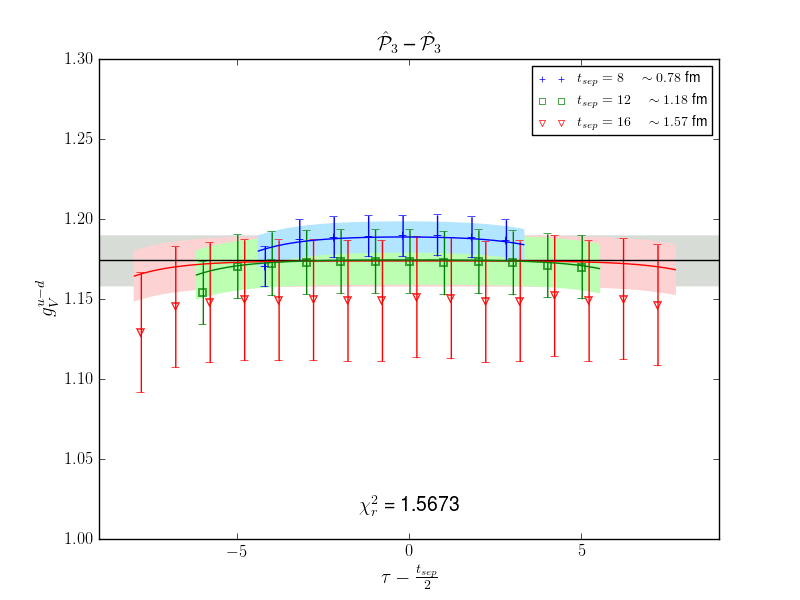}
    \includegraphics[width=0.49\linewidth]{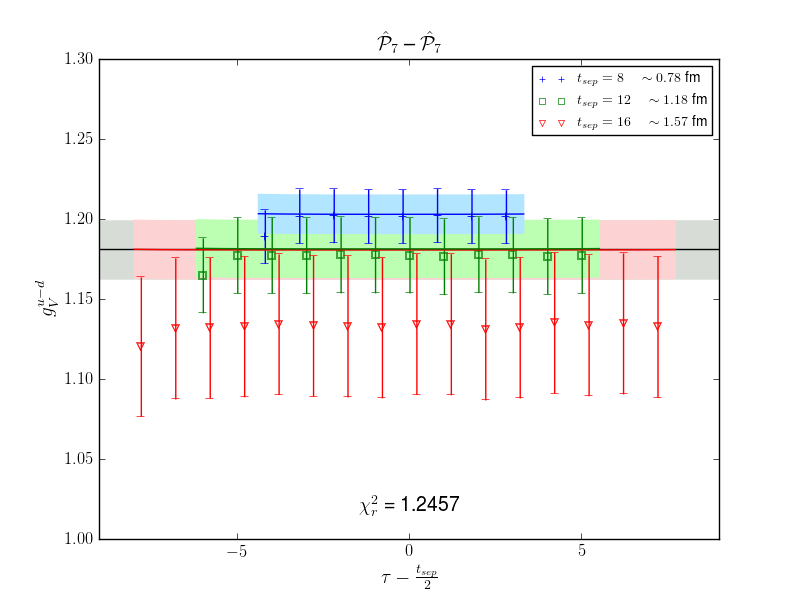}
  \caption{Extracted effective unrenormalized isovector vector charge using the projected distilled interpolator from the 3x3 GEVP (left) and the projected distilled interpolator from the 7x7 GEVP (right). Displayed plots are for simultaneous fits with $t^{\text{fit}}_{2\text{pt}}\in\left[2,16\right]$ and $\tau_{\text{buff}}=2$.}
\end{figure*}
\begin{table*}[tb]
  \centering
  \scriptsize{
    \begin{tabular}{ | c | c | c | c | c | c | c | c | c | c |}
      \hline
      $\hat{\mathcal{O}}$ & $\tau_{\text{buff}}$ & $\mathcal{A}$ & $\mathcal{B}$ & $\mathcal{C}$ & $M_0$ & $M_1$ & $\left|\mathbf{a}\right|^2$ & $\left|\mathbf{b}\right|^2$ & $g^{u-d}_{V,\text{bare}}$ \\
      \hline
      \multirow{4}{*}{Jacobi-SS} & 1 & 4.68(31)e-08 & 8.30(9.65)e-08 & -8.62(1.37)e-10 & 0.543(6) & 1.049(77) & 4.13(23)e-08 & 3.73(25)e-08 & 1.131(47) \\
      & 2 & 4.70(32)e-08 & 8.34(9.55)e-08 & -9.62(3.94)e-10 & 0.543(6) & 1.047(77) & 4.13(23)e-08 & 3.71(25)e-08 & 1.137(47) \\
      & 3 & 4.71(32)e-08 & 8.26(9.28)e-08 & 0.07(1.57)e-09 & 0.543(6) & 1.043(77) & 4.13(23)e-08 & 3.69(24)e-08 & 1.140(48) \\
      & 4 & 4.97(97)e-08 & 3.19(8.09)e-07 & -0.52(1.75)e-07 & 0.543(6) & 1.043(76) & 4.12(23)e-08 & 3.70(24)e-08 & 1.21(23) \\ \hline
      \multirow{4}{*}{$\thinspace^2S_S\frac{1}{2}^+$} & 1 & 1.73(03)e-02 & 4.90(5.49)e-02 & -1.33(0.13)e-04 & 0.536(1) & 1.255(29) & 1.46(02)e-02 & 1.69(06)e-02 & 1.185(14) \\
      & 2 & 1.73(03)e-02 & 4.82(5.31)e-02 & -8.84(3.77)e-05 & 0.536(1) & 1.251(29) & 1.46(02)e-02 & 1.69(06)e-02 & 1.184(14) \\
      & 3 & 1.73(03)e-02 & 4.90(5.31)e-02 & 2.52(1.63)e-04 & 0.536(1) & 1.251(29) & 1.46(02)e-02 & 1.69(06)e-02 & 1.184(14) \\
      & 4 & 1.65(05)e-02 & -4.01(2.63)e-01 & 3.99(2.17)e-02 & 0.536(1) & 1.249(29) & 1.46(02)e-02 & 1.69(06)e-02 & 1.131(32) \\ \hline
      \multirow{4}{*}{$\hat{\mathcal{P}}_3$} & 1 & 1.25(2)e+00 & 9.64(7.17)e+00 & -1.60(0.11)e-02 & 0.536(2) & 1.306(39) & 1.06(1)e+00 & 1.19(6)e+00 & 1.174(15) \\
      & 2 & 1.25(2)e+00 & 9.56(7.55)e+00 & -1.74(0.34)e-02 & 0.536(2) & 1.310(40) & 1.07(1)e+00 & 1.19(6)e+00 & 1.174(16) \\
      & 3 & 1.26(2)e+00 & 9.00(7.61)e+00 & -0.21(1.54)e-02 & 0.536(2) & 1.315(41) & 1.07(1)e+00 & 1.20(6)e+00 & 1.177(15) \\
      & 4 & 1.18(8)e+00 & -65.7(75.8)e+00 & 4.99(4.97)e+00 & 0.536(2) & 1.313(41) & 1.07(1)e+00 & 1.20(6)e+00 & 1.107(72) \\ \hline
      \multirow{4}{*}{$\hat{\mathcal{P}}_7$} & 1 & 1.18(2)e+00 & 27.9(19.8)e+00 & -1.67(1.16)e-03 & 0.535(2) & 1.412(70) & 9.98(13)e-01 & 1.05(11)e+00 & 1.180(17) \\
      & 2 & 1.18(2)e+00 & 26.0(23.2)e+00 & 0.68(3.75)e-03 & 0.535(2) & 1.412(75) & 9.98(13)e-01 & 1.05(11)e+00 & 1.181(18) \\
      & 3 & 1.18(3)e+00 & 26.1(23.4)e+00 & 1.82(1.94)e-02 & 0.535(2) & 1.418(75) & 9.99(13)e-01 & 1.06(11)e+00 & 1.183(18) \\
      & 4 & 1.07(6)e+00 & -294(237)e+00 & 13.1(7.1)e+00 & 0.535(2) & 1.427(77) & 9.99(13)e-01 & 1.08(12)e+00 & 1.068(53) \\ \hline
    \end{tabular}
  }
  \caption{Results of simultaneous fits to the two-point and three-point correlators with vector current insertion. The functional forms are given in Eqns. \ref{eq:2ptfit} and \ref{eq:3ptfit}.}
\end{table*}

\subsection{Numerical Cost}
Given the substantial benefits incurred by use of distillation and the variational method in such calculations of hadronic structure, it is worth while to pause and consider the numerical cost of doing so. We highlight that a true one-to-one comparison of Distillation to standard techniques is not entirely possible, as distillation is markedly distinct from traditional spatial smearing techniques via sampling of entire time slices.

The calculation of point-to-all propagators in standard lattice calculations proceeds by inverting the chosen discretization of the Dirac operator against a point source in coordinate, spinor, and color space
\begin{equation}
  S\left(\vec{x},\vec{z}\right)^{\beta\alpha}_{ba}=\sum_{\vec{y},\gamma,c}D^{-1}\left(\vec{x},\vec{y}\right)^{\beta\gamma}_{bc}\delta\left(\vec{y}-\vec{z}\right)\delta_{\gamma\alpha}\delta_{ca}.
\end{equation}
This operation requires 12 distinct inversions of the Dirac operator, one for each spinor and color index $\lbrace\alpha,a\rbrace$. In our case of the nucleon, with degenerate $u$ and $d$-quarks, this captures the propagation of the nucleon from some source point to all other points on the lattice. Implementing the sequential-source method, as we did for the Jacobi-SS interpolator, reduces the number of required inversions by combining the computed point-to-all propagator with the sink interpolator and using this object as a source for further inversions (deemed sequential propagators). As the $u$-quarks can be combined into one sequential source and the $d$-quarks another, the computation of a three-point function in the sequential-source framework requires two additional inversions against a color and spinor point source. As the sequential propagators must be recomputed for each new source-sink separation $t_{\text{sep}}$, we arrive at
\begin{equation}
  N_{\text{src}}\left[12+24\cdot N_{\text{seps}}\right]N_{\text{cfg}}
\end{equation}
total inversions of the Dirac operator for a single Jacobi-smeared interpolator, where $N_{\text{src}}$ is the number of source positions, $N_{\text{seps}}$ the number of source-sink separations, and $N_{\text{cfg}}$ the number of gauge configurations. Were the variational method applied to a two-point correlation matrix of different Jacobi-smeared interpolators, $12\times N_{\text{ops}}$ inversions would be required to construct the requisite variationally-improved interpolator. The total number of inversions in the three-point function remains unchanged, and we would then have
\begin{equation}
  N_{\text{src}}\left[12\left(1+N_{\text{ops}}\right)+24\cdot N_{\text{seps}}\right]N_{\text{cfg}}
\end{equation}
inversions, with $N_{\text{ops}}$ the dimension of the interpolator basis. If instead the variational method were applied to a correlation matrix of different Jacobi-smeared three-point functions, $N_{\text{ops}}$ inversions would be needed at the source and an additional $N_{\text{ops}}$ in the construction of the sequential propagators, for a total
\begin{equation}
  N_{\text{src}}N_{\text{ops}}\left[12+24\cdot N_{\text{seps}}N_{\text{ops}}\right]N_{\text{cfg}},
\end{equation}
where we note that this is now proportional to $N_{\rm ops}^2$.

In the case of distillation, the inversion of the Dirac operator against a point source is replaced with inversion against an eigenvector on some time slice $t$:
\begin{equation}
  S_{\alpha\beta}^{\left(k\right)}\left(\vec{x},t';t\right)=D^{-1}_{\alpha\beta}\left(t',t\right)\xi^{\left(k\right)}\left(t\right).
\end{equation}
As the eigenvectors are determined by solution of
$-\nabla^2\left(t\right)\xi^{\left(k\right)}\left(t\right)=\lambda^{\left(k\right)}\left(t\right)\xi^{\left(k\right)}\left(t\right)$
given some gauge covariant discretization of $\nabla^2$, the
calculation of the $k^{\text{th}}$ solution vector requires 3
inversions of the Dirac operator, one for each (suppressed) color
index. In practice, for a given $t_{\text{sep}}$, we calculate the
solution vectors forward (backward) from the source (sink), where the
solution vectors from the source are used in the two-point and
three-point calculations. Thus for a single distilled interpolator the
total number of inversions becomes
\begin{equation}
  3\cdot N_{\text{src}}N_{\text{eigs}}\left(1+N_{\text{seps}}\right)N_{\text{cfg}}
\end{equation}
where $N_{\text{eigs}}$ is the dimension of the distillation
space. With $N_{\text{eigs}}=64$ in our case, this cost at first seems
excessive. However, once these solution vectors have been computed,
any number of interpolating fields, variationally-optimized or
otherwise, can be correlated \textit{without} additional cost. We note
that we have not taken into account the cost of the Wick comparisons
when using distillation in this study, and a future work will include
the stability of our extracted matrix elements as the rank of the
distillation space is reduced.
  
\subsection{Conclusions}
We have investigated the use of distillation, and an extended basis of
interpolators, in the calculation of the scalar, axial and tensor
isovector charges of the nucleon, and made comparisons with a
calculation on the same ensemble using the conventional
Jacobi-smearing method of a single smearing radius.  We find that
distillation affords a considerable improvement in the statistical
quality of the data when compared with calculations using Jacobi
smearing. We attribute this improvement to be
a consequence of momentum projection performed at both source
and sink in the case of a two-point function, and at source, sink and
current in the case of a three-point function.  More surprisingly,
even the use of a single, local distilled interpolating operator results in a
suppression of the contribution of excited-states relative to that of
the ground-state in both two-point and three-point functions.

For our variational analysis, we employed a basis of operators
comprising the non-relativistic operators that can be constructed with
up to two covariant derivatives, together with so-called ``hybrid''
operators where the gluons play a manifest role.  Whilst the
improvement was not as dramatic as that between a single
Jacobi-smeared and a single distillation-smeared operator, the use of
the variational method and the extended basis provided more consistency and fidelity in the matrix elements for different source-sink separations.
Furthermore, the extended basis can be introduced without further
propagator calculations, in contrast to the case of Jacobi smearing
where the use of the variational method demands a considerable
increase in the number of propagators to be computed.

The next step in our program is to extend our investigation from
matrix elements between nucleons at rest to those in motion, and from
forward matrix elements to off-forward matrix elements.  The former is
key to the efficient application of the quasi-PDF, pseudo-PDF and
current-current correlator methods to the calculation of parton distribution functions in the
nucleon, whilst the latter is important for form factors at high
momenta, and off-forward matrix elements such as generalized parton
distributions.  Nonetheless, the work presented here clearly
demonstrates the efficacy of distillation as a means both of
decreasing the statistical uncertainty, and of reducing excited-state
contributions, in calculations of nucleon properties.

\subsection*{Acknowledgments}
Calculations were performed using the \textit{Chroma} \cite{Edwards:2004sx}
software suite on the computing clusters at Jefferson Lab.
We are grateful to Jo Dudek and Stefan Meinel for the use
of their fitting codes, and to Robert Edwards for invaluable discussions
on the feasibility of these calculations. CE extends thanks to
Balint Jo\'o for invaluable discussions on building software on the
varied machine architectures on the Jefferson Lab clusters. This
material is based upon work supported by the U.S. Department of
Energy, Office of Science, Office of Nuclear Physics under contract
DE-AC05-06OR23177. Computations for this work were carried out in part
on facilities of the USQCD Collaboration, which are funded by the
Office of Science of the U.S.\ Department of Energy.  CE was supported
in part by the U.S. Department of Energy under contract
DE-FG02-04ER41302 and a Department of Energy Office of Science
Graduate Student Research fellowship, through the U.S. Department of
Energy, Office of Science, Office of Workforce Development for
Teachers and Scientists, Office of Science Graduate Student Research
(SCGSR) program. The SCGSR program is administered by the Oak Ridge
Institute for Science and Education (ORISE) for the DOE. ORISE is
managed by ORAU under contract number DE-SC0014664.

\bibliography{charge_srcs}
\end{document}